\documentclass[fleqn,usenatbib]{mnras}

\usepackage{newtxtext,newtxmath}

\usepackage[T1]{fontenc}
\usepackage{ae,aecompl}

\usepackage{graphicx}	
\usepackage{amsmath}	
\usepackage{amssymb}	

\usepackage{multirow} 
\usepackage[referable]{threeparttablex} 
\usepackage{subfig} 
\usepackage{booktabs} 

\title[System IMF of 25 Ori]{System IMF of the 25 Ori Group from Planetary-Mass Objects to Intermediate/High-Mass Stars}

\author[G. Su\'arez et al.]{
Genaro Su\'arez,$^{1}$\thanks{E-mail: gsuarez@astro.unam.mx}
Juan Jos\'e~Downes,$^{2,3}$
Carlos Rom\'an-Z\'u\~niga,$^{1}$
Miguel Cervi\~no,$^{4,5,6}$
\newauthor
C\'esar~Brice\~no,$^{7}$
Monika G.~Petr-Gotzens$^{8}$
\& Katherina Vivas$^{7}$
\\
$^{1}$Instituto de Astronom\'ia, Universidad Nacional Aut\'onoma de M\'exico, Unidad Acad\'emica en Ensenada, Ensenada BC 22860, M\'exico\\
$^{2}$Centro Universitario Regional del Este, Universidad de la Rep\'ublica, AP 264, Rocha 27000, Uruguay\\
$^{3}$Centro de Investigaciones de Astronom\'ia, Apartado Postal 264, M\'erida, Venezuela \\
$^{4}$Centro de Astrobiolog\'ia (CSIC/INTA), 28850 Torrej\'on de Ardoz, Madrid, Spain\\
$^{5}$Instituto de Astrof\'{i}sica de Canarias, c/v\'{i}a L\'actea s/n, 38205 La Laguna, Tenerife, Spain\\
$^{6}$Instituto de Astrof\'{i}sica de Andaluc\'ia (CSIC), 18008, Granada, Spain\\
$^{7}$Cerro Tololo Interamerican Observatory, Casilla 603, La Serena, Chile\\
$^{8}$European Southern Observatory, Karl-Schwarzschild-Str. 2, 85748 Garching bei M\"unchen, Germany
}

\date{Accepted XXX. Received YYY; in original form ZZZ}

\pubyear{2015}

\begin{document}
\label{firstpage}
\pagerange{\pageref{firstpage}--\pageref{lastpage}}
\maketitle

\begin{abstract}
The stellar initial mass function (IMF) is an essential input for many astrophysical studies but only in a few cases it has been determined over the whole cluster mass range, limiting the conclusions about its nature. The 25 Orionis group (25 Ori) is an excellent laboratory to investigate the IMF across the entire mass range of the population, from planetary-mass objects to intermediate/high-mass stars. We combine new deep optical photometry with optical and near-infrared data from the literature to select 1687 member candidates covering a 1.1$^\circ$ radius area in 25 Ori. With this sample we derived the 25 Ori system IMF from 0.012 to 13.1 $M_\odot$. This system IMF is well described by a two-segment power-law with $\Gamma=-0.74\pm0.04$ for $m<0.4\ M_\odot$ and $\Gamma=1.50\pm0.11$ for $m\ge0.4\ M_\odot$. It is also well described over the whole mass range by a tapered power-law function with $\Gamma=1.10\pm0.09$, $m_p=0.31\pm0.03$ and $\beta=2.11\pm0.09$. The best lognormal representation of the system IMF has $m_c=0.31\pm0.04$ and $\sigma=0.46\pm0.05$ for $m<1\ M_\odot$. This system IMF does not present significant variations with the radii. We compared the resultant system IMF as well as the BD/star ratio of $0.16\pm0.03$ we estimated for 25 Ori with that of other stellar regions with diverse conditions and found no significant discrepancies. These results support the idea that general star formation mechanisms are probably not strongly dependent to environmental conditions. We found that the substellar and stellar objects in 25 Ori have similar spatial distributions and confirmed that 25 Ori is a gravitationally unbound stellar association.
\end{abstract}

\begin{keywords}
brown dwarf - stars: luminosity function, mass function - stars: low-mass - stars: formation - stars: pre-main sequence - open clusters and associations: individual: 25 Orionis
\end{keywords}

\section{Introduction}
\label{sec:introduction}
The mass spectrum of the members of a stellar population at birth is known as initial mass function (IMF). The IMF is the main product of the star formation process and is one of the fundamental astrophysical quantities. Since the seminal IMF study by \citet{Salpeter1955}, there have been many contributions to this topic to understand the origin and behaviour of the IMF, but only few of them focus on the whole mass range of the populations, which limits the conclusions about its complete shape \cite[e.g.][and references therein]{Bastian2010}.

Observational IMF studies in a complete range of masses, from planetary-mass objects to massive star scales, allow to analyse the continuity of the star formation process over about three orders of magnitude of mass and help to constrain initial conditions of star formation models. These kind of studies are also important to understand if the star formation process is sensitive or not to environmental conditions and if it changes in time, which is the nature of the so-called \textit{universality of the IMF} \cite[e.g.][]{Kroupa2013,Offner2014}.

Young stellar clusters ($\lesssim 10$ Myr) are useful laboratories for observational studies of the IMF in a wide range of masses because objects are brighter in the pre-main sequence (PMS) phase than on the main sequence (MS), none or minimum correction by the stellar evolution of their members is necessary, their spatial distributions are relatively small (for groups beyond the solar neighbourhood) and their members have basically the same age, metallicity and distance. However, an important issue to be taken into account when working with embedded clusters \citep[$\lesssim3$ Myr; ][]{Lada-Lada2003} is dust extinction, which, on one hand, complicates the detection of the least massive objects but, on the other hand, helps to separate the cluster population from the background contamination. An additional issue that can affect IMF determinations is the loss of low-mass members in dynamically evolved clusters caused by the preferential escape of these members and/or by the brown dwarf (BD) photospheric cooling \citep{deLaFuenteMarcos-deLaFuenteMarcos2000}. Therefore, we should look for stellar clusters that are old enough to diminish extinction effects but are also young enough to allow a complete determination of the IMF.

The best studied clusters in the literature in terms of their IMFs over a wide mass range are: Pleiades \citep[0.03~-~10 $M_\odot$;][]{Moraux2003}, Blanco 1 \citep[0.03~-~3 $M_\odot$;][]{Moraux2007a}, $\sigma$ Ori \citep[0.006~-~19 $M_\odot$;][]{PenaRamirez2012}, the Orion Nebula Cluster \citep[ONC; 0.025~-~3 $M_\odot$, $\approx$0.005~-~1 $M_\odot$;][respectively]{DaRio2012,Drass2016}, and RCW 38 \citep[0.02~-~20 $M_\odot$;][]{Muzic2017}. Additionally to these studies based on photometric data, a detailed determination of the IMF of Collinder 69 based on spectroscopically confirmed members across more than three orders of magnitude of mass (0.016~-~20 $M_\odot$) was presented by \citet[]{Bayo2011}. All these studies reported the IMF not corrected by unresolved multiple systems, also referred as {\it system} IMF \citep{Chabrier2003a}. Additionally, \citet{Moraux2003} and \citet{Muzic2017} also presented the single-star IMF, in which a correction by multiple systems is applied. In Table \ref{tab:imf_literature} we summarize the resulting parametrizations of these system IMFs as well as the employed theoretical models for mass determination. For parametrizations of a larger sample of clusters but in smaller mass ranges see Table 1 from \citet{DeMarchi2010} and Table 4 from \citet{Muzic2017}, mainly, for low-mass stars (LMSs). Although the tables indicated above show some differences between the various IMFs, more complete and systematic observational studies are needed in populations with different environments and evolutionary stages before any claim concerning variations of the IMF, as suggested by \citet{Bastian2010} and \citet{Offner2014}.

\begin{table*} \scriptsize
 \caption{System IMF parametrizations over a wide mass range in several young clusters.}
 \label{tab:imf_literature}
 \begin{threeparttable}
  \begin{tabular}{@{\extracolsep{2pt}}lcccccccccc@{}}
   \toprule
 	Cluster      & Age   &       \multicolumn{3}{c}{Lognormal}          &         \multicolumn{4}{c}{Power Law}                & Model  & Ref  \\
   \cline{3-5}
   \cline{6-9}
 	             &       & $m_c$         &    $\sigma$   & $m$ range    & $\Gamma_1^a$ & $m$ range   & $\Gamma_2^b$  & $m$ range   &        & \\
 	             & [Myr] & [$M_\odot$]   &               & [$M_\odot$]  &              & [$M_\odot$] &               & [$M_\odot$] &        & \\
   \midrule
   \multirow{2}{*} {RCW 38} & \multirow{2}{*} {1$^c$} &               &               &              & -0.29$\pm$0.11     & 0.02-0.50  & 0.60$\pm$0.13     & 0.50-20 & \multirow{2}{*} {BT-Settl+PARSEC} & \multirow{2}{*} {1} \\
                            &       &               &               &              & -0.58$\pm$0.18     & 0.02-0.20  & 0.48$\pm$0.08     & 0.20-20 &                                   &                     \\
   \multirow{2}{*} {ONC} & \multirow{2}{*} {2} & 0.35$\pm$0.02$^d$ & 0.44$\pm$0.05$^d$ & \multirow{2}{*} {0.025-3} & -1.12$\pm$0.90$^d$     & 0.025-0.30 & 0.60$\pm$0.33$^d$     & 0.30-3  & NextGen & \multirow{2}{*} {2} \\
                         &   & 0.28$\pm$0.02 & 0.38$\pm$0.01     &                           & -2.41$\pm$0.25     & 0.025-0.17 & 1.30$\pm$0.09     & 0.17-3  & DM98 &                                    \\
   \multirow{2}{*} {$\sigma$ Ori} & \multirow{2}{*} {$\sim$3$^e$} & 0.24$\pm$0.09$^f$ & 0.53$\pm$0.19$^f$ & 0.006-1  & \multirow{2}{*} {-0.45$\pm$0.20} & \multirow{2}{*} {0.006-0.35} & \multirow{2}{*} {0.70$\pm$0.20} & \multirow{2}{*} {0.35-19} & \multirow{2}{*} {Siess+Lyon} & \multirow{2}{*} {3} \\
                                  &       & 0.27$\pm$0.09$^f$ & 0.63$\pm$0.15$^f$ & 0.006-19 &                                  &                              &                                 &                           &                                  &                                           \\
   Collinder 69 & 5$^g$     &              &               &              & -0.71$\pm$0.10$^h$ & 0.01-0.65  & 0.82$\pm$0.05$^h$ & 0.65-25 & Siess+COND    & 4 \\
   Blanco 1     & 100-150     &0.36$\pm$0.07 & 0.58$\pm$0.06 & 0.03-3       & -0.31$\pm$0.15     & 0.03-0.60  &                   &         & NextGen+DUSTY & 5 \\
   Pleiades     & 125$^i$     & 0.25          & 0.52          & 0.03-10      & -0.40$\pm$0.11     & 0.03-0.48  & 1.7               & 1.5-10  & NextGen       & 6 \\
   \bottomrule
  \end{tabular}
  \begin{tablenotes}[para,flushleft]
    $^a$For LMSs and BDs.\\
    $^b$For intermediate/high-mass stars.\\
    $^c$\citet{Getman2014b}.\\
    $^d$For sources older than 1 Myr.\\
    $^e$\citet{ZapateroOsorio2002} and \citet{Caballero2008b}.\\
    $^f$Mean values of the two set of parameters obtained combining \citet{Baraffe1998} and \citet{Siess2000} models at different cutoffs (0.3 and 1 $M_\odot$).\\
    $^g$\citet{Dolan1999} and \citet{Bayo2011}.\\
    $^h$Mean value of the six reported values and the error as the standard deviation.\\
    $^i$\citet{Stauffer1998}.\\

    NextGen: \citet{Baraffe1998}, DM98: \citet{DAntona1998}, Siess: \citet{Siess2000}, DUSTY: \citep{Chabrier2000}, COND: \citep{Baraffe2003}, Lyon: NextGen, DUSTY and COND, BT-Settl: \citet{Baraffe2015}, and PARSEC: \citet{Bressan2012} and \citet{Chen2014}.\\

    References: (1) \citet{Muzic2017}, (2) \citet{DaRio2012}, (3) \citet{PenaRamirez2012}, (4) \citet{Bayo2011}, (5) \citet{Moraux2007a}, and (6) \citet{Moraux2003}.
  \end{tablenotes}
 \end{threeparttable}
\end{table*}

An interesting young stellar group for studying the IMF over its whole mass range and full spatial extent is 25 Orionis (25 Ori), the most prominent spatial overdensity of PMS stars in Orion OB1a, originally detected by \citet{Briceno2005} and kinematically confirmed by \citet{Briceno2007}. The estimated area of this group have radii of 1.0$^\circ$ \citep{Briceno2005,Briceno2007}, 0.5$^\circ$ \citep[hereafter referred as the 25 Ori overdensity; ][]{Downes2014} and 0.7$^\circ$ \citep{Briceno2019}, centred at $\alpha_{J2000}=81.2^\circ$ and $\delta_{J2000}=1.7^\circ$. This makes it feasible to perform an observational study covering the full spatial extent of this group. 25 Ori is a 7-10 Myr population located at 356$\pm$47 pc and presents a low visual extinction of 0.29$\pm$0.26 mag (see Appendix \ref{sec_app:25Ori_params}), which facilitates the detection of members down to planetary-masses \citep{Downes2015}. Several previous studies have focused on characterizing the 25 Ori population; \citet{Kharchenko2005,Kharchenko2013} for intermediate/high-mass stars, \citet{Briceno2005,McGehee2006,Briceno2007,Hernandez2007a,Biazzo2011b,Downes2014,Suarez2017,Briceno2019} for LMSs and \citet{Downes2015} for BDs.

In 2014, \citeauthor{Downes2014} reported the first and only available determination of  the system IMF of 25 Ori in the mass range $0.02\lesssim m/M_\odot\lesssim 0.8$ working with a sample of photometric member candidates inside an area of 3x3 deg$^2$ around 25 Ori. In this work we improve the 25 Ori system IMF by including optical and near-infrared (NIR) photometry spanning from intermediate/high-mass stars down to planetary-mass objects ($0.012\le m/M_\odot\le13.1$) and also covering its full spatial extent. In Section \ref{sec:data}, we present our observations and public catalogues used in this study. The selection of the photometric member candidates and a discussion of different issues that could affect the determination of the IMF, in the particular case of 25 Ori, and how we deal with them are presented in Section \ref{sec:candidates}. In Section \ref{sec:results}, we present the derivation of the 25 Ori system IMF and the comparisons with other associations, and the analysis of the spatial distribution, BD frequency and gravitational state of 25 Ori. Finally, a summary and conclusions are given in Section \ref{sec:conclusions}.

\section{Photometric Data}
\label{sec:data}

\subsection{DECam observations}
\label{sec:DECam}
\begin{figure*}
	\includegraphics[width=0.85\textwidth]{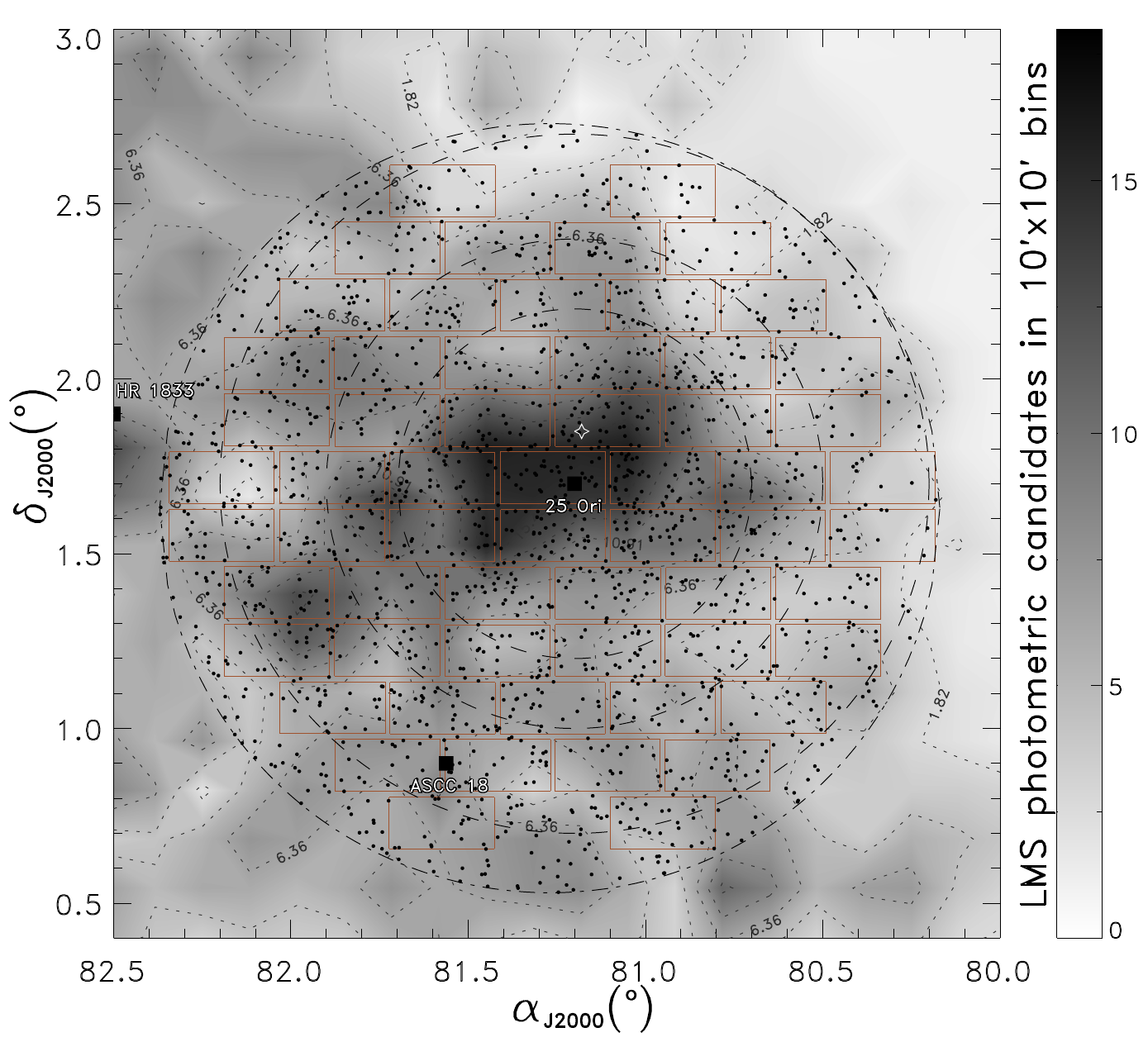}
	\caption{Spatial distribution of our photometric member candidates (black points; see Section \ref{sec:candidates}). The dash-dotted circle shows the FOV of our DECam observations obtained with the array of detectors indicated by the brown boxes. The dashed circles indicate, from the centre outwards, the 25 Ori estimated areas by \citet[0.5$^\circ$ radius; ][]{Downes2014}, \citet[0.7$^\circ$ radius; ][]{Briceno2019} and \citet[1.0$^\circ$ radius; ][]{Briceno2005,Briceno2007} centred at $\alpha_{J2000}=81.2^\circ$ and $\delta_{J2000}=1.7^\circ$. The black squares indicate the labelled stellar groups (25 Ori by \citealt{Briceno2005}, ASCC 18 by \citealt{Kharchenko2013} and HR 1833 by \citealt{Briceno2019}). The grey background map indicates the density of LMS and BD photometric member candidates of Orion OB1a, computed as the number of sources in bins of 10'x10' using the selection of \citet{Downes2014}. The white star symbol shows the position of the 25 Ori star.}
	\label{fig:sky}
\end{figure*}

This work includes new very deep optical \textit{i}-band photometry of 25 Ori obtained using the Dark Energy Camera (DECam) mounted on the 4m Victor M. Blanco telescope at CTIO. DECam is a 570 Megapixel camera with an array of 62 2kx4k detectors with a plate scale of $0.263''$ pixel$^{-1}$, covering a field of view (FOV) of 1.1$^\circ$ radius \citep{Flaugher2015}. Our DECam observations were performed on Feb 24, 2016 (PI: G. Su\'arez). We obtained 11x300s exposures in the $i$-band centred at $\alpha_{J2000}= 05^{\rm h} 25^{\rm m} 04^{\rm s}.8$ and $\delta_{J2000} = +01^{\circ} 37' 48''.6$ with an airmass $<1.3$ and a mean seeing of $\sim 0.9''$. During our observations two DECam detectors were not functional, reducing the array to 60 usable detectors. In Section \ref{sec:spatial} we discuss how this fact, together with the gaps and the non circular configuration of the detectors, affect the spatial coverage of the DECam observations. In Figure \ref{fig:sky} we show the spatial coverage of our DECam data.

The reduced and calibrated data were produced by the DECam Community Pipeline \citep{Valdes2014} and downloaded from the NOAO Science Archive\footnote{\url{http://archive.noao.edu/}}. The resulting data have processing level of 2, which means they are single reduced frames after removing the instrument signature and applying the WCS and photometric calibrations, as explained in the NOAO Data Handbook\footnote{\url{http://ast.noao.edu/sites/default/files/NOAO\_DHB\_v2.2.pdf}}.

We combined the individual frames using the \texttt{imcombine} routine of IRAF\footnote{IRAF is distributed by NOAO, which is operated by AURA, Inc., under cooperative agreement with the NSF.}, considering a ccdclip value of 3.5-$\sigma$ and correcting the offset of the individual images using the WCS solutions provided by the NOAO pipeline. The accuracy of the astrometry with respect to the the Sloan Digital Sky Survey (SDSS) Data Release 9 \citep{Ahn2012} catalogue is 0.1$^{\prime\prime}$ for bright sources and decays to 0.2$^{\prime\prime}$ for the faintest sources in the DECam catalogue. The photometry was made using a modification of the $PinkPack$ pipeline \citep{Levine2006} to work with the DECam data, which uses the SExtractor software \citep{Bertin-Arnouts1996} for the detections, IRAF/APPHOT for the aperture photometry and IRAF/DAOPHOT for the PSF photometry. To calibrate the resulting $i$-band photometry we added the zero point of 25.18 mag for our DECam observations and an offset of 0.637 mag with respect to the $i$-band photometry in the DECam system obtained from the SDSS catalogue. More details about this calibration are found in Appendix \ref{sec_app:DECam_calibration}. The mean value of the residuals between our calibrated data and those in the DECam system using photometry from SDSS is -0.001 mag with a RMS of 0.038 mag.

\subsection{CIDA Deep Survey of Orion}
\label{sec:CDSO}
Additional optical $I_c$-band photometry for sources brighter that the DECam saturation limit (see Section \ref{sec:sensitivity}) was obtained from the CIDA Deep Survey of Orion \citep[CDSO; ][]{Downes2014}. This catalogue was constructed by coading the photometry from the CIDA Variability Survey of Orion \citep[CVSO; ][]{Briceno2005,Mateu2012,Briceno2019}, obtained at the National Astronomical Observatory of Venezuela. The area covered by this survey extends beyond the limits of our DECam data.

\subsection{VISTA Orion Survey}
\label{sec:VISTA}
The deep $Z$, $J$ and $K$ near-infrared photometry for this study is from the VISTA survey in Orion \citep{Petr-Gotzens2011}, which was carried out as part of the VISTA science verification program \citep{Arnaboldi2010} with the near-infrared camera (VIRCAM) mounted on the 4.2m telescope at Paranal Observatory.

\subsection{Photometry from Literature}
\label{sec:catalogues}
\subsubsection{Optical Photometry}
The optical data from DECam and the CDSO were complemented with the $i$-band photometry from the UCAC4 catalogue \citep{Zacharias2013} as well as the $I_c$-band photometry from the $Hipparcos$ catalogue \citep{Perryman1997} for the brightest sources in 25 Ori.

\subsubsection{Near-IR Photometry}
We complemented the VISTA near-infrared photometry with $J$ and $Ks$-band photometry from the 2MASS catalogue \citep{Skrutskie2006}.

In Table \ref{tab:catalogues} we summarized the spatial coverage of 25 Ori (for an area of 0.7$^\circ$ radius, see Section \ref{sec:spatial}), the spatial resolution and the photometric sensitivities (see Section \ref{sec:sensitivity}) of the optical and NIR catalogues used in this study. The masses corresponding to the saturation and completeness magnitudes are obtained using the mass-luminosity relation explained in Section \ref{sec:mass-luminosity}.

\begin{table*}
\caption{Spatial coverage of 25 Ori$^a$ and photometric sensitivities of the catalogues used in this study.}
  \small
  \label{tab:catalogues}
  \begin{threeparttable}
    \setlength{\tabcolsep}{10pt}
 	\begin{tabular}{lcccccccc}
    \toprule
 	Survey      & Phot.    & FWHM      & Area         & Satur.       & Comp.        & Satur.     & Comp.       &  Ref. \\
 	            & Band     & (arcsec)  & (per cent)         & (mag)        & (mag)        & ($M_\odot$)& ($M_\odot$) & \\
    \midrule
 	DECam       & $I_c$    & 0.9       & $\approx 86$ & 16.0         & 22.50        & 0.16       & 0.012       & a     \\
 	CDSO        & $I_c$    & 2.9       & 100          & 13.0         & 19.75        & 0.86       & 0.020       & b     \\
 	UCAC4       & $I_c$	   & 1.9       & 100          & 7.0          & 14.75        & 6.33       & 0.340       & c     \\
 	$Hipparcos$ & $I_c$	   & ---       & 100          & $<$5.0       & ---          & $>$13.5    & ---         & d     \\
 	VISTA       & $J$      & 0.9       & 100          & 12.0         & 20.25        & 0.85       & $<$0.010    & e     \\
 	2MASS       & $J$      & 2.5       & 100          & 4.0          & 16.25        & 19.3       & 0.287       & f     \\
    \bottomrule
 	\end{tabular}
  \begin{tablenotes}[para,flushleft]
    $^a$Considering an area of $0.7^\circ$ radius.\\
	References: (a) This work; (b) \citet{Downes2014}; (c) \citet{Zacharias2013}; (d) \citet{Perryman1997}; (e) \citet{Petr-Gotzens2011}; (d) \citet{Skrutskie2006}
  \end{tablenotes}
 \end{threeparttable}
\end{table*}

\subsection{Merged Optical-NIR Catalog}
\label{sec:merged_cat}

From the individual catalogues with optical and NIR data we constructed one single general catalogue, as explained in this section.

\subsubsection{Transformation of optical photometry into Cousins system}
\label{sec:photometry_transformation}
We transformed the $i$-band photometry from UCAC4 and DECam to the Cousin system $I_c$-band, which is a photometric band predicted by the BT-Settl \citep{Baraffe2015} and PARSEC-COLIBRI \citep{Marigo2017} isochrones used to estimate masses to later construct the system IMF in Section \ref{sec:mass-luminosity}. To obtain the $I_c$ magnitudes from UCAC4 we used the empirical transformations by \citet{Jordi2006}, which relate SDSS photometry with other photometric systems included the Cousins system. For the DECam photometry we derived directly from our data colour-dependent transformations to convert the calibrated DECam magnitudes to the SDSS system and then to the Cousins system. The RMS we obtained when comparing the $I_c$ magnitudes from the CDSO and those from UCAC4 and from DECam after the transformation are 0.07 and 0.04 mag, respectively. The details about these transformations are described in Appendix \ref{sec_app:photometry_transformation}. Because the Cousins photometric system is already used by the CDSO and $Hipparcos$ catalogues, after the transformation of the DECam and UCAC4 photometries, the complete sample of optical observations are all in the same photometric system.

\subsubsection{Photometric uncertainties}
\label{sec:photometry_uncertainties}
\begin{figure}
	\includegraphics[width=0.47\textwidth]{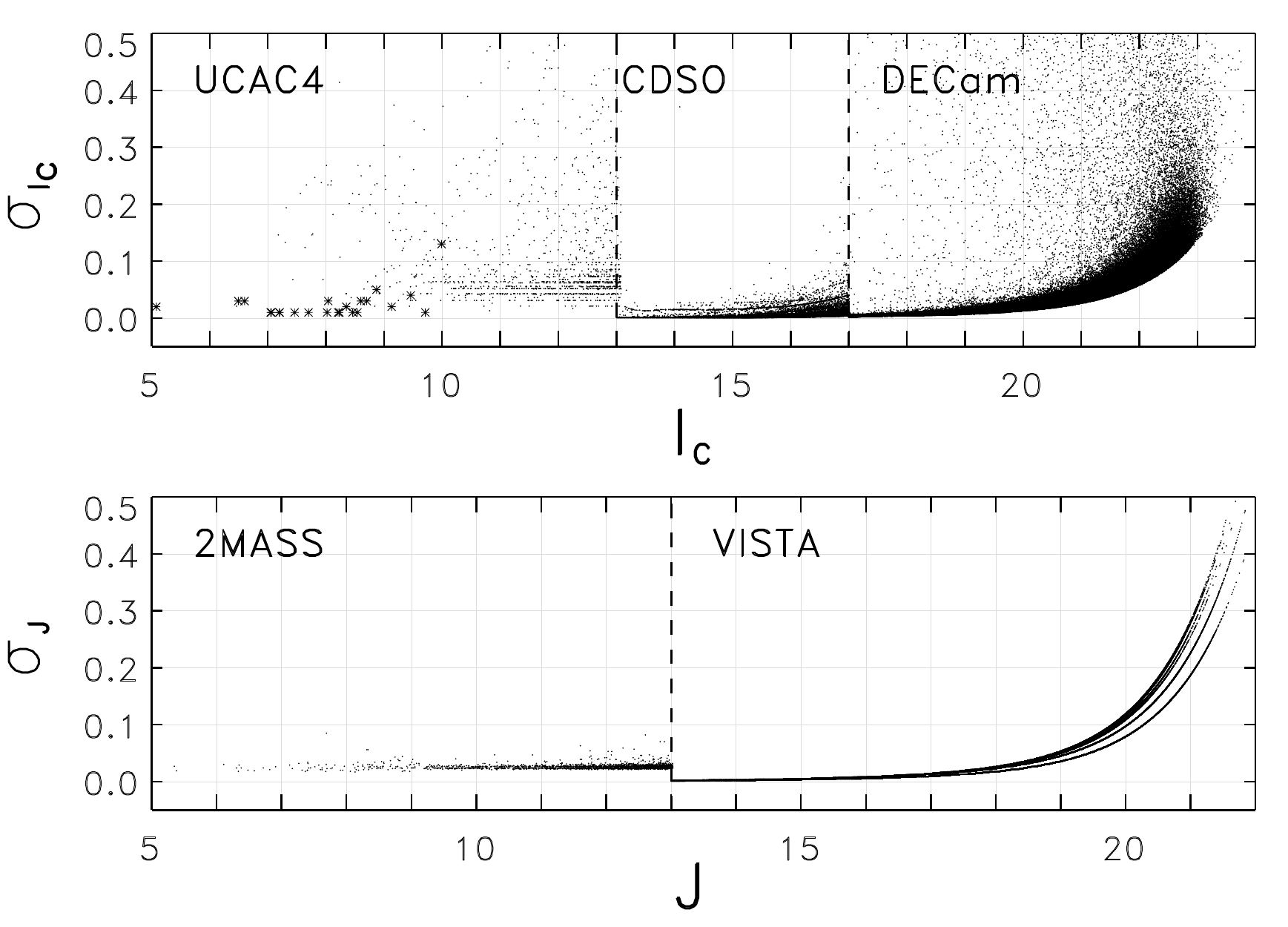}
	\caption{Photometric uncertainties as a function of magnitude for the merged optical (top) and NIR (bottom) catalogues. The labelled names indicate the catalogues used in each magnitude range separated by the dashed lines. The few $Hipparcos$ sources in the optical catalogue are indicated by the asterisks.}
	\label{fig:errors}
\end{figure}

Before we define the brightness ranges where each catalogue will be used, we fitted exponential functions to the photometric uncertainties of the optical and NIR catalogues with respect to the magnitude ($\delta I_c(I_c)$ for the optical data and $\delta J(J)$ for the NIR data). This way we can estimate the uncertainties of the data as a function of the photometric magnitudes, which will allow us to combine the catalogues considering the typical photometric uncertainties at each brightness point where the catalogues are joined. In Figure \ref{fig:errors} we show the photometric uncertainties of the catalogues we used and in Table \ref{tab:errors} we list the parameters of the functions fitted to these uncertainties working with magnitudes inside their saturation and completeness limits (see Section \ref{sec:sensitivity}).

\begin{table}
\caption{Parameters of the exponentials fitted to the photometric uncertainties of the optical and NIR catalogues used in this study.}
  \small
  \label{tab:errors}
  \begin{threeparttable}
 	\begin{tabular}{lcccc}
    \toprule
 	Catalog & Photometric &  $a$  &  $b$   & $c$   \\
			& Band        &       &        &       \\
    \midrule
 	DECam   & $I_c$	& 0.005 & 25.861 & 1.042 \\
 	CDSO    & $I_c$	& 0.002 & 22.175 & 0.999 \\
 	UCAC4   & $I_c$	& 0.037 & 8.993  & 0.453 \\
 	VISTA   & $J$	& 0.002 & 16.870 & 0.732 \\
 	2MASS   & $J$	& 0.024 & 20.240 & 1.105 \\
    \bottomrule
 	\end{tabular}
  \begin{tablenotes}[para,flushleft]
	Note. The exponentials have the form $f(x)=a+e^{(cx-b)}$, where $x$ is the magnitude in the corresponding photometric band.
  \end{tablenotes}
 \end{threeparttable}
\end{table}

\subsubsection{Cutoffs and merged catalogues}
\label{sec:cutoffs}
The brightness ranges where each photometric catalogue was used are related to their photometric sensitivities, which are described in Section \ref{sec:sensitivity} and reported in Table \ref{tab:catalogues}. The $I_c$-band photometry we used to have a combined optical catalogue are as follows: $i)$ UCAC4 for $I_c<13.0+\delta I_c(13.0)$, $ii)$ CDSO for $13.0-\delta I_c(13.0)\leq I_c<17.0+\delta I_c(17.0)$, and $iii)$ DECam for $I_c\geq 17.0-\delta I_c(17.0)$. We also added 25 stars (including 25 Ori) from the $Hipparcos$ catalogue, which are too bright to have $I_c$ magnitudes from UCAC4. The $J$-band photometry used to have a combined NIR catalogue are as follows: $i)$ 2MASS for $J<13.0+\delta J(13.0)$, and $ii)$ VISTA for $J\geq 13.0-\delta J(13.0)$. Then, we removed 3$^{\prime\prime}$ duplicates from the optical and NIR catalogues and kept the sources with smaller photometric uncertainties. To join the optical and NIR catalogues we did a cross-match between them with a tolerance of 3$^{\prime\prime}$ using STILTS\footnote{\url{http://www.star.bris.ac.uk/~mbt/stilts/}} \citep{Taylor2006}.

The final optical and NIR catalogue has 110527 detections inside an area of 1.1$^\circ$ radius in 25 Ori, being most of them (about 85 per cent) from the DECam and VISTA catalogues.

\section{Selection of Photometric Candidates}
\label{sec:candidates}

\subsection{PMS Locus}
\label{sec:locus}
\begin{figure*}
	\includegraphics[width=0.85\textwidth]{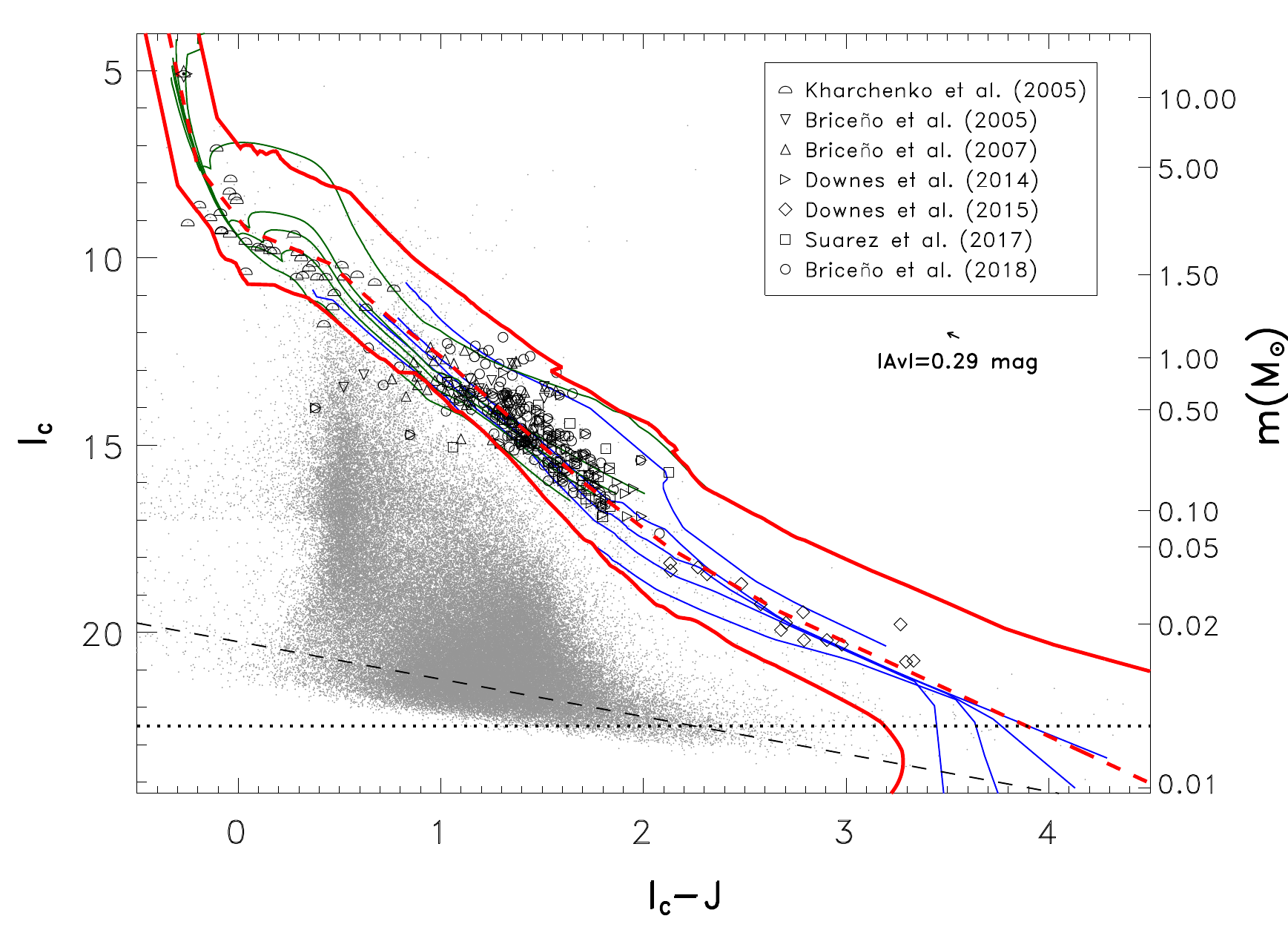}
	\caption{CMD used for the selection of photometric member candidates of 25 Ori. The red solid curves show the PMS locus defined considering the empirical isochrone (red dashed curve) and several issues that may affect the position of the sources in this plot. The open symbols represent the known spectroscopically confirmed members \citep{Briceno2005,Briceno2007,Downes2014,Downes2015,Suarez2017,Briceno2019} and high-probable members \citep{Kharchenko2005} of 25 Ori, as shown in the label, which trace the empirical isochrone. The grey dots indicate all the detections in our combined optical and NIR catalogue. The black dotted and black dashed lines show the $I_c$/DECam and $J$/VISTA completeness magnitudes, respectively. The blue and green curves indicate, respectively, the BT-Settl and PARSEC-COLIBRI isochrones for ages, from top to bottom, of 1, 5, 7, 10 and 20 Myr. The arrow shows the dereddening vector for the mean extinction of 25 Ori. The right axis indicates the corresponding masses from the PARSEC-COLIBRI 7 Myr isochrone for $m>1\ M_\odot$ and from the BT-Settl 7 Myr isochrone for lower masses, considering a distance of 356 pc and a visual extinction of 0.29 mag. The giants and subgiants branches cross the PMS locus close to (0.9, 13) and (0.5, 11), respectively.}
	\label{fig:CMD}
\end{figure*}

The use of colour-magnitude diagrams (CMDs) combining optical and NIR data has been successfully tested for identifying young stellar objects \citep[e.g. ][ and references therein]{Downes2014}. We selected photometric member candidates from the merged optical and NIR catalogue according to their position in the $I_c$ vs $I_c-J$ digram shown in Figure \ref{fig:CMD}.

To define the PMS locus in which the member candidates lie, we plotted a large set of 355 spectroscopically confirmed low-mass members of 25 Ori from \cite{Briceno2005,Briceno2007,Downes2014,Suarez2017,Briceno2019} and 15 spectroscopically confirmed BD members of 25 Ori and Orion OB1a from \citet{Downes2015}. Most of these members were confirmed through similar spectroscopic procedures, which makes the sample more homogeneous. Additionally to the confirmed members, we also plotted 38 highly probable intermediate/high-mass members from \cite{Kharchenko2005}. The final sample of 408 spectroscopically confirmed members and highly probable members covers the spectral type range from B2 to M9 and trace a clear sequence in the $I_c$ vs $I_c-J$ diagram. This sequence corresponds to the empirical isochrone of 25 Ori, which was defined averaging the $I_c-J$ colours per $I_c$-bin (red dashed curve in Figure \ref{fig:CMD}). The resulting empirical isochrone is roughly consistent with the PARSEC-COLIBRI and BT-Settl 7 Myr isochrones, confirming the 25 Ori age \citep[6.1$\pm$2.4; ][ and references therein]{Briceno2019}. This empirical isochrone was our starting point to define the PMS locus considering the following uncertainties and effects:

$i)$ \emph{Distance uncertainty.} From the sample of spectroscopically confirmed members of 25 Ori by \citet{Briceno2005,Briceno2007,Downes2014,Downes2015,Suarez2017,Briceno2019}, we obtained a mean distance of 356 pc with a standard deviation, $\sigma$, of 47 pc, considering the distance estimates we calculated from the Gaia parallaxes \citep[Gaia DR2; ][]{GaiaCollaboration2018} with uncertainties of $\le 20$ per cent using the method implemented by \citet{Bailer-Jones2015} and \citet[BJ18; ][]{Bailer-Jones2018}, as explained in Appendix \ref{sec_app:distance}. Then, we broaden vertically the edges of the PMS locus in the CMD by adding the 1-$\sigma$ uncertainty in distance, which corresponds to upwards and downwards offsets of 0.31 and 0.27 mag, respectively.

$ii)$ \emph{Age uncertainty.} To estimate the change in the $I_c$ brightness ($\Delta I_c$) as a function of the $I_c-J$ colour due to the uncertainty of the 25 Ori age \citep[6.1$\pm$2.4 Myr; ][ and references therein]{Briceno2019}, we worked with the PARSEC-COLIBRI and BT-Settl isochrones. We obtained $\Delta I_c^I$ between the isochrone corresponding to the age of 25 Ori and that for the 25 Ori age minus the error. Similarly, we obtained $\Delta I_c^{II}$ considering the age of 25 Ori and the age plus the error. In most of the colour range considered (-0.5-4.5 mag), $\Delta I_c^I$ is larger than $\Delta I_c^{II}$. We used $\Delta I_c^I$ to move upwards the upper edge of the locus and $\Delta I_c^{II}$ to move downwards the lower edge.

$iii)$ \emph{Unresolved binarity.} According to \citet{Briceno2007}, the observed spread in the CMD of young stars in the 25 Ori field is roughly consistent with the upper limit of 0.75 mag expected from unresolved binaries. Thus, we used this limit to move upward the upper edge of the locus.

$iv)$ \emph{Mean intrinsic variability.} We characterized the $I_c$-amplitude variations as a function of the magnitude for the 25 Ori member candidates from \citet{Downes2014} using the CVSO catalogue. These amplitude variations increase with the $I_c$ magnitudes from 0.2 to 0.9 mag in the brightness range between 13.0 to 19.0 mag. For brighter and fainter $I_c$ magnitudes we assumed these minimum and maximum variation limits, respectively. Thus, we used these $I_c$-amplitude variations to move upwards and downwards the upper and lower edges of the locus, respectively. For the $J$-band, \citet{Scholz2009} reported the low-level amplitude variations of about 0.2 mag for young LMSs and BDs. Assuming that when occurs a maximum or a minimum in the $I_c$-brightness of a variable source also takes place the maximum or minimum in the $J$-band brightness, we considered $I_c-J$ amplitude variations as the difference between the $I_c$-amplitude variations and the representative 0.2 mag variations in the $J$-band to move leftwards and rightwards the blue (lower) and red (upper) edges of the locus, respectively.

$v)$ \emph{Photometric uncertainties.} We considered the exponentials fitted to the uncertainties of the optical and NIR catalogues as a function of the magnitudes to move both edges of the locus. The upper and lower edges were moved upwards and downwards, respectively, according to the uncertainty corresponding to each $I_c$-magnitude of the optical catalogues used in the different ranges. The blue (lower) and red (upper) edges of the locus were moved leftwards and rightwards, respectively, considering the uncertainties added in quadrature for each $I_c$ and $J$-magnitude from the catalogues used in the different ranges.

The sources lying inside this resulting PMS locus were selected as photometric member candidates of 25 Ori. We selected 1694 candidates inside the DECam FOV having $I_c$ magnitudes from 5.08 to 23.3 mag. In Table \ref{tab:candidates_catalogue} we provide the list of our member candidates together with the optical and NIR photometry used in this study after removing potential extragalactic sources in Section \ref{sec:extgalcontam}. We also included in this table the corresponding citation for the cases when a candidate has been studied in previous contributions.

The locus defined this way contains about 95 per cent of the confirmed members and highly probable members of 25 Ori. From the members lying out, on the left side, of the PMS locus, about 75 per cent of them have $>99$ per cent probability of being variable stars in the CVSO. In Section \ref{sec:missed} we estimated that the fraction of 25 Ori members we can lose in our photometric selection is $\sim3.1$ per cent.

It is important to notice in Figure \ref{fig:CMD} that in the $I_c$ range roughly between 9 and 13 mag, the giant and subgiant branches cross the PMS locus, which increases the contamination by these sources in this brightness range. We discussed in Section \ref{sec:fieldcontam} how to deal with this contamination.

\begin{table*}
\caption{List of photometric member candidates used in this study.}
  \small
  \label{tab:candidates_catalogue}
  \begin{threeparttable}
    \setlength{\tabcolsep}{4pt}
 	\begin{tabular}{@{\extracolsep{0pt}}lccccccccccccc@{}}
    \toprule
	ID & $\alpha_{J2000}$ & $\delta_{J2000}$ & $I_c$ & e$I_c$ & $J$ & e$J$ & $H$ & e$H$ & $K$ & e$K$ & Source $I_c$ & Source $JHK$ & Ref \\

    \midrule
  	25Ori\_1    & 81.186774 &  1.846467 &  5.08   &  0.02  &  5.349  &  0.027 &  5.417  &  0.027 &  5.355  &  0.02  &  Hipparcos &  2MASS  &      b   \\ 
  	25Ori\_20   & 81.208733 &  0.766291 &  8.569  &  0.217 &  8.761  &  0.026 &  8.805  &  0.055 &  8.826  &  0.019 &  UCAC4     &  2MASS  &      c   \\ 
  	25Ori\_734  & 81.166427 &  1.361305 &  13.384 &  0.001 &  12.278 &  0.023 &  11.609 &  0.022 &  11.436 &  0.026 &  CDSO      &  2MASS  &      g   \\ 
  	25Ori\_898  & 81.664442 &  1.20085  &  14.309 &  0.001 &  13.008 &  0.002 &  12.399 &  0.002 &  12.187 &  0.002 &  CDSO      &  VISTA  &      k   \\ 
  	25Ori\_1642 & 81.243645 &  1.733347 &  20.154 &  0.032 &  17.358 &  0.012 &  16.762 &  0.012 &  16.241 &  0.025 &  DECam     &  VISTA  &      i   \\ 
  	25Ori\_1685 & 80.573212 &  1.435463 &  22.602 &  0.137 &  19.257 &  0.055 &  18.442 &  0.06  &  18.108 &  0.075 &  DECam     &  VISTA  &      p   \\ 
    \bottomrule
 	\end{tabular}
  \begin{tablenotes}[para,flushleft]
	References: \\
		(a) \citet{Briceno2005}; spectroscopically confirmed members. \\
		(b) \citet{Hernandez2005}; member candidates using kinematic and photometric data. \\
		(c) \citet{Kharchenko2005}; highly probable candidates. \\ 
		(d) \citet{Kharchenko2005}; low-probability candidates. \\ 
		(e) \citet{Briceno2007}; spectroscopically confirmed members. \\
		(f) \citet{Hernandez2007a}; member candidates using infrared and optical photometric data. \\
		(g) \citet{Downes2014}; spectroscopically confirmed members. \\
		(h) \citet{Downes2014}; photometric member candidates using optical and NIR data. \\
		(i) \citet{Downes2015}; spectroscopically confirmed members. \\
		(j) \citet{Downes2015}; sources rejected as members. \\
		(k) \citet{Suarez2017}; spectroscopically confirmed members. \\
		(l) \citet{Suarez2017}; sources rejected as members. \\
		(m) \citet{Kounkel2018}; highly problable members using kinematic data. \\
		(n) \citet{Kounkel2018}; candidates from \citet{Cottle2018} rejected as members. \\
		(o) \citet{Briceno2019}; spectroscopically confirmed members. \\
		(p) This work. \\ \\
	(This table is available in its entirety in machine-readable form.)
  \end{tablenotes}
 \end{threeparttable}
\end{table*}

\subsection{Sources of Uncertainty, Contamination and Biases}
\label{sec:uncertainties}
Several previous works have studied the uncertainties and biases implicit in the observational determination of the IMF \citep[e.g.][]{Moraux2003,Moraux2007a,Moraux2007b,Ascenso2011,Bayo2011,Jeffries2012,Dib2017}. In this section we characterize these effects in the case of 25 Ori and show how we corrected them.

\subsubsection{Spatial Completeness}
\label{sec:spatial}
The CDSO and VISTA catalogues and all the public catalogues considered in this work have a full spatial coverage of the FOV of the DECam observations.

As explained in Section \ref{sec:DECam}, our DECam observations were obtained with an array of 60 detectors configured as shown in Figure \ref{fig:sky} (brown boxes), therefore, part of the area in a FOV is lost by the gaps and because the array is not circular. To compute what fraction of a FOV is covered by the DECam data, we used the Monte Carlo method to generate a list of sources randomly distributed inside the FOV and counting those lying inside the detectors. We found this way that for the DECam FOV, the DECam data cover $\approx70$ per cent of the area. If we consider the previously estimated areas of 25 Ori, the DECam observations have a coverage of $\approx 79$ per cent when considering \citet[1.0$^\circ$ radius; ][]{Briceno2005,Briceno2007} and $\approx86$ per cent when considering \citet[0.7$^\circ$ radius; ][]{Briceno2019} or \citet[0.5$^\circ$ radius; ][]{Downes2014}. These fractions will allow us to correct the luminosity function (LF) and system IMF of 25 Ori by the spatial coverage of the DECam data considering that the LMSs and BDs in 25 Ori do not present any preferential spatial distribution (see Section \ref{sec:spatial_distribution}). In Table \ref{tab:catalogues} we report the spatial coverage of 25 Ori for all the catalogues used in this study.

In Table \ref{tab:candidates} we list the number of member candidates inside the DECam FOV after applying the correction by the spatial coverage of the DECam data. If we had a full coverage of the DECam observations, we would expect 1782 photometric member candidates in the $I_c$ range from 5.08 to 23.3 mag. The mass range corresponding to this brightness range is obtained in Section \ref{sec:sys-imf}.

\subsubsection{Photometric Sensitivity}
\label{sec:sensitivity}
The saturation and completeness magnitudes for the optical and NIR catalogues were determined, respectively, as the brightest and faintest magnitudes between which the logarithmic number of sources per magnitude bin do not deviate from a linear behaviour. We estimated the masses corresponding to these magnitudes using the PARSEC-COLIBRI 7 Myr isochrone for $m>1\ M_\odot$ and the BT-Settl 7 Myr isochrone for lower masses. In Table \ref{tab:catalogues} we summarize these values, where we can see how the optical and NIR catalogues complement each other. Therefore, in the determination of the LF and system IMF of 25 Ori, for the sources more massive than the DECam completeness mass (0.012 $M_{Jup}$), it is not necessary to make any correction due to the photometric sensitivity of the catalogues.

\subsubsection{Contamination by Field Stars}
\label{sec:fieldcontam}
Though the use of optical-NIR CMDs allows a clear selection of young sources, a contamination of $\sim20$ per cent by field stars is expected for the low-mass domain \citep{Downes2014} and $\sim30$ per cent for the very low-mass and BD regime \citep{Downes2015} in our sample of photometric member candidates. Furthermore, a higher degree of contamination is expected in the intermediate-mass range of our candidate sample due to giant and subgiant stars.

We estimated the number of field stars inside the PMS locus following two procedures: First, by means of a simulation of the expected galactic stellar population using the Besan\c{c}on Galactic model \citep[hereafter BGM; ][]{Robin2003}. Second, empirically, by a fiducial selection of photometric candidates from an observed control field with similar galactic latitude.

For the BGM approach we performed four simulations\footnote{\url{http://model2016.obs-besancon.fr}} in an area of 2x2 deg$^2$ in 25 Ori and considering the photometric uncertainties of our joined optical and NIR catalogues shown in Figure \ref{fig:errors} and listed in Table \ref{tab:errors}. The simulated populations combined the optical and NIR photometric errors from UCAC4 and 2MASS (simulation 1), CDSO and 2MASS (simulation 2), CDSO and VISTA (simulation 3), and DECam and VISTA (simulation 4). Then, we joined the resulting simulations by keeping the sources brighter than $I_c=13$ mag from simulation 1, the sources in the range 13 mag$\le I_c<$15 mag from simulation 2, the sources with magnitudes 15 mag$\le I_c<$17 mag from simulation 3, and sources with $I_c \ge 17$ mag from simulation 4. This way we have a simulated stellar population compatible with our observational joined optical-NIR catalogue.

For the control field approach, we estimated the field star contamination in our candidate sample by means of direct counting on selected regions as follows: $i)$ For the optical CDSO, UCAC4 and $Hipparcos$, and NIR 2MASS catalogues, we considered a control field of $1.0^\circ$ radius FOV placed at the same galactic latitude of 25 Ori in a direction moving away from the Orion's Belt ($\alpha_{J2000}= 05^{\rm h} 19^{\rm m} 03^{\rm s}.6$ and $\delta_{J2000} = +04^{\circ} 18' 17''.1$). $ii)$ Since we do not have neither DECam nor VISTA specific observations in this region, we used for these catalogues the areas of the eight north-westernmost and westernmost detectors of the DECam array as control fields, because a) they mostly lie outside the larger estimated area of 25 Ori, b) they have the lesser number of Orion OB1a reported members \citep{Briceno2019,Kounkel2018} and c) the density of LMS and BD candidates in the regions covered by these detectors falls to about 10 per cent of the density in the 25 Ori core \citep{Downes2014}. Then, we joined all the photometric catalogues from both control fields in the same way we did for the 25 Ori observations.

We applied our procedure for selecting photometric member candidates to the BGM and control field samples in order to account the sources lying inside the PMS locus, which we defined as contaminants. The number of contaminants in both samples are consistent for magnitudes brighter than $I_c\sim 17$ mag, as discussed in Section \ref{sec:LF}. In Table \ref{tab:candidates} we list the number of member candidates and contaminants after applying the spatial coverage corrections for the DECam data as well as their complete brightness and mass ranges. Using the control field we estimated that the fraction of contaminants present in our candidate sample, in the $I_c$ brightness range between 13 and 20 mag, is about 30 per cent, which is somewhat higher than the 20 per cent estimated, and spectroscopically proven, by \citet{Downes2014} in the same brightness range for their candidate selection working with similar CMDs but using a narrower PMS locus. In Section \ref{sec:sample} we compared both samples. 

As mentioned in Section \ref{sec:locus} and shown in Figure \ref{fig:CMD}, there is a high contamination by giant and subgiant stars in the $I_c$ range between $\sim$9 and $\sim$13 mag in our candidate sample. Even, the contaminants estimated by the control field or the BGM can be as numerous as the member candidates in this particular brightness range, which do not allow us to remove the contamination in this range using only the control field or BGM. Fortunately, we can take advantage of Gaia DR2 because about 92 per cent of the candidates brighter than $I_c=13$ mag have parallaxes with errors of $\le 20$ per cent. The same fraction is obtained for candidates with brightness up to about 17.5 mag, but for fainter sources Gaia DR2 starts presenting significant incompleteness issues. Thus, we did a subset of the member candidates with $I_c<17.5$ mag and having distances and proper motions within 3-$\sigma$ of the mean values of 25 Ori (356$\pm$47 pc and $\mu_\alpha = 1.33\pm0.46$ mas yr$^{-1}$ and $\mu_\delta = -0.23\pm0.55$ mas yr$^{-1}$; see Appendix \ref{sec_app:25Ori_params}). Additionally, we removed the sources with deviant radial velocities \citep[$<$15 km s$^{-1}$ or $>$40 km s$^{-1}$; ][]{Briceno2007} from \citet{Kounkel2018}. With these criteria we recover about 90 per cent of the confirmed members. Hereafter, we are going to refer to this subset of highly probable 25 Ori members as the filtered sample of candidates.

From the member candidates in the $I_c$ brightness range between 9 and 13 mag, only $\approx$11 per cent of them satisfy the distance and proper motion criteria. About 70 per cent of the sources contaminating this brightness range and having parallax errors of $\le 20$ per cent have distances significantly larger than those of the 25 Ori members. Thus, we checked that these contaminants are, in fact, giant or subgiant stars, as predicted by the BGM.

Field stars are the main, but not the only, contamination present in our candidate sample. After applying the correction for the DECam spatial coverage, we have only one BGM contaminant fainter than $I_c=19.6$ mag, while there are about 32 contaminants using the control field in the same brightness range. As the BGM does not include extragalactic sources, this difference between the contaminants counted in both samples suggests that most of the contamination present in the faintest range of our candidate sample is due to extragalactic sources. 

\begin{table}
\caption{Number, $I_c$ brightness and mass ranges of the member candidates and contaminants in an area of 1.1$^\circ$ radius in 25 Ori after correcting by the spatial coverage of the DECam data.}
	\small
	\label{tab:candidates}
 	\begin{tabular}{@{}lccc}
    \toprule
 	Origin  	       & Number            & $I_c$ Range  & Mass Range \\ \vspace{-.05in}
					   & of Sources        &              &            \\
 	    	   		   & 	        	   & (mag)        & ($M_\odot$)\\
    \midrule
 	25 Ori FOV         & 1782	  		   & 5.08-23.3    & 0.011-13.1   \\
 	Control Field FOV  & 1030 	  		   & 6.51-23.3    & 0.011-7.74     \\ 
 	BGM                & 840   	  		   & 7.67-19.6$^a$& 0.021$^a$-4.76 \\ 
    \bottomrule
 	\end{tabular}
	$^a$There is a fainter dwarf star contaminant with $I_c=23.5$ (0.011 $M_\odot$).
\end{table}

\subsubsection{Contamination by Extragalactic Sources}
\label{sec:extgalcontam}
\begin{figure}
	\includegraphics[width=0.47\textwidth]{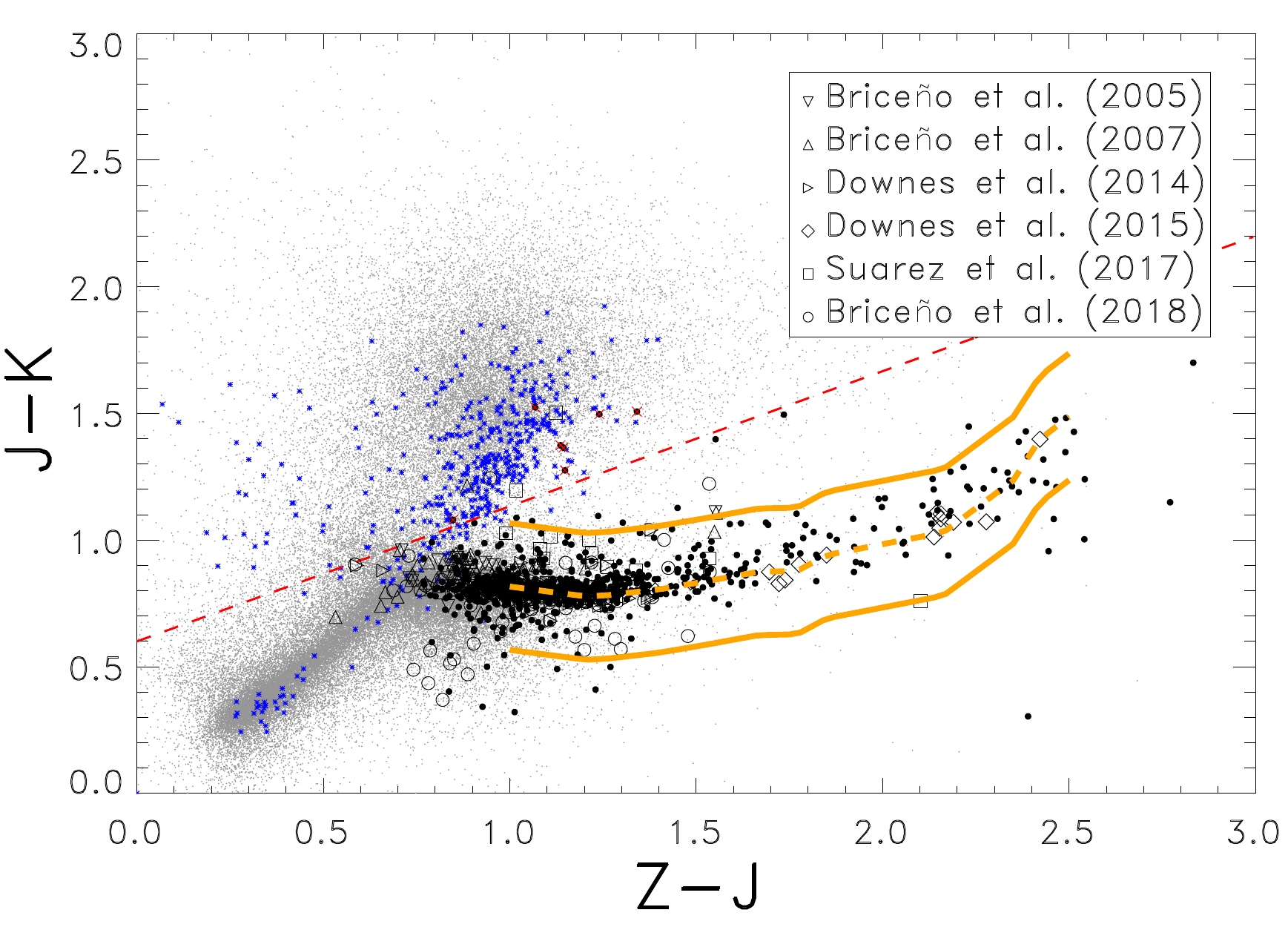}
	\caption{Colour-colour diagram used to remove highly probable extragalactic contaminants (red crosses) from our member candidate sample (black dots). The blue asterisks represent a sample of spectroscopically confirmed galaxies and quasars in the direction of 25 Ori \citep{Suarez2017}. The red dashed line separates more than 80 per cent of the sample of extragalactic sources from the member candidates. The orange dashed curve shows the empirical isochrone traced by the low-mass and BD confirmed members of 25 Ori by the studies indicated in the label, which are mostly contained in the sequence defined by the orange solid curves. The grey dots are the same as in Figure \ref{fig:CMD}.}
	\label{fig:CCD}
\end{figure}

As 25 Ori is out of the galactic plane ($b=18.4^\circ$) and has a low visual extinction of 0.29$\pm$0.26 mag, we expect extragalactic sources in any deep photometric sample in that direction. We suggest in the previous section that the contamination by extragalactic sources dominates the contamination in the faintest range of our member candidate sample. To remove the most likely extragalactic sources from this sample we used the $J-K$ vs $Z-J$ colour-colour diagram shown in Figure \ref{fig:CCD}. We plotted a sample of $\approx 500$ spectroscopically confirmed galaxies and quasars in the direction of 25 Ori with $I_c$-brightness between 13.5 and 20.0 mag from \cite{Suarez2017}. Also, we plotted our member candidates and the previously confirmed members of 25 Ori. Similarly to what we did for the CMD, we defined the empirical isochrone traced by the low-mass and BD confirmed members. Then, we defined the sequence centred on this isochrone and containing over 90 per cent of the confirmed members. This sequence is clearly distinct from the region where are located more than $80$ per cent of the galaxies and quasars. About $1$ per cent (7 sources) of the member candidates plotted in this colour-colour diagram (those having VISTA photometry) lie in the region defined by the galaxy/quasar sample and have $I_c$ magnitudes between 15.2 and 18.2 mag. We considered these 7 sources as contaminants and removed them from our member candidate sample, keeping the rest of the candidates selected in the CMD. The resultant sample has 1687 member candidates and is provided in Table \ref{tab:candidates_catalogue}. This is the list of candidates we used to derive the LF and system IMF of 25 Ori in Sections \ref{sec:LF} and \ref{sec:IMF}, respectively.

We used a similar colour-colour diagram to that in Figure \ref{fig:CCD} to remove potential extragalactic sources present in the contaminants from the control field. 

In Figure \ref{fig:CCD}, only four ($\sim$1 per cent) of the spectroscopically confirmed members lie in the region where most of the galaxies and quasars are located. Two of these peculiar members are classical T-Tauri stars (CTTSs) harbouring circumstellar discs and having an intense H$_\alpha$ emission \citep[41 and 53 \AA; ][]{Suarez2017}, while the other two have low H$_\alpha$ emission, one being a CTTS and the other one a weak T-Tauri star \citep[WTTS; ][]{Briceno2007}. These four members are highly probable to be variable stars according to the CVSO, which could explain their position in the colour-colour diagram.

After we removed from our member candidate sample and from the control field contaminants the potential extragalactic sources, we used the control field to statistically remove the extragalactic and galactic contamination from the LF and system IMF of 25 Ori in Section \ref{sec:LF} and \ref{sec:IMF}.

\subsubsection{IR excesses}
\label{sec:excesses}
Possible excesses in the $J$-band, due to discs, can bias the candidate selection because members showing such excesses could lie outside, on the red side, of the PMS locus. In Figure \ref{fig:CMD} there are 60 sources lying on the right side of the PMS locus, which have $I_c<12.7$ mag. The simulations performed with the BGM show that the positions of these sources are consistent with those of giant stars. Additionally, we checked the distances of these sources and compared them with the currently estimated 25 Ori distance. 97 per cent of the sources having parallaxes with errors of $\le 20$ per cent have distances not consistent with those of the 25 Ori members, of which most of them (91 per cent) have larger distances, suggesting these are, in fact, giant stars. Only two sources have distances consistent with 25 Ori, but these sources have unexpected photometric uncertainties from the UCAC4 catalogue (0.146 and 0.234 mag), which could explain, in part, their location in the CMD. Thus, most of the sources left out, on the red side, of the PMS locus are behind the 25 Ori population, indicating that in our photometric selection we do not lose 25 Ori members due to the presence of IR excesses.

Additionally, if the magnitudes used to obtain the masses were affected by the IR excesses, the masses could be overestimated. However, at the age of 25 Ori, only a fraction of $\sim$5 per cent of the LMSs harbour circumstellar discs \citep{Briceno2005,Briceno2007,Hernandez2007a,Downes2014,Briceno2019}, which produce IR excesses starting at the $WISE\ 3.4\ \mu$m band or longer wavelengths \citep{Suarez2017}. Even for the BDs in 25 Ori, which have a larger disc fraction of $\sim 30$ per cent, the IR excesses start beyond the $K$-band \citep{Downes2015}. In this study we used the $I_c$ and $J$-band magnitudes which are not expected to be affected by IR excesses. In any event, we worked with the $I_c$ magnitudes to estimate masses to avoid any overestimation due to IR excesses.

\subsubsection{Effects of Chromospheric Activity}
\label{sec:activity}
Active LMSs suppress the effective temperature by $\sim5$ per cent and inflate the radius by $\sim10$ per cent with respect to inactive objects \citep[e.g. ][]{Lopez-Morales2007}. These effects roughly cancel themselves, which preserves the bolometric luminosity \citep{Stassun2012}. 

Due to the effective temperature suppression, the masses of active LMSs estimated from the H-R diagram are underestimated, but if masses are estimated from luminosities (or absolute magnitudes), the effect would be much smaller \citep{Jeffries2017}. According to \citet{Stassun2012}, when the effective temperature is used to estimate masses from model isochrones, the resultant masses are systematically lower than the true masses by factors of $\sim3$ and $\sim2$ for LMSs and BDs with intense chromospheric activity of $\log L_{H_\alpha}/L_{bol}=-3.3$, respectively. This level of chromospheric activity corresponds to the saturation limit in young LMSs, which separates the CTTSs from WTTSs \citep{BarradoYNavascues-Martin2003}. For LMSs and BDs with low levels of magnetic activity ($\log L_{H_\alpha}/L_{bol}=-4.5$), the masses estimated using the effective temperature are systematically lower than true values by factors of $\sim2$ and $\sim1.5$, respectively. Instead, when masses are estimated using bolometric luminosities derived from $K$-band absolute magnitudes and considering model isochrones, the resulting masses are $\sim5$ per cent smaller than true values for LMSs and BDs with high chromospheric activity and roughly unaffected for LMSs and BDs with low chromospheric activity \citep{Stassun2012}. This small bias is introduced by the effective temperature dependence of the bolometric corrections that \citet{Stassun2012} used to convert absolute magnitudes to bolometric luminosities.

In our case, as explained in Section \ref{sec:mass-luminosity}, we used absolute magnitudes and model isochrones to obtain the masses of the member candidates which, according to \citet{Stassun2012}, minimize the underestimation bias of masses of active stars introduced when the bolometric luminosities of the models are transformed to absolute magnitudes. Additionally, the fraction of active stars in 25 Ori is $\sim 5$ per cent \citep[][]{Briceno2005,Briceno2007,Hernandez2007a,Downes2014,Briceno2019}. Considering the expected $\sim$5 per cent underestimation of masses for the expected $\sim$5 per cent of active stars in our candidate sample, we estimated that the change in the system IMF of 25 Ori is smaller than the Poisson noise of the distribution.

\subsubsection{Spatial Resolution and Binaries}
Most of the mass distributions of stellar clusters available in the literature do not take into account unresolved binaries or multiple systems and are, in fact, the system IMFs \citep[e.g. Table \ref{tab:imf_literature} and ][]{Bastian2010}.

A revision and treatment of the effect of unresolved binary systems in the IMF parametrization is found in \citet{Muzic2017}. They found that the mass distribution becomes steeper in the low-mass and high-mass sides when correcting the system IMF by binary systems to obtain the single-star IMF, but the changes in the slopes agree within the uncertainties. A similar effect on the IMF due to binary systems is reported in \citet{Kroupa2001b}.

In this study we reported the system IMF of 25 Ori, which will allow us to directly compare it with all the system IMFs in Table \ref{tab:imf_literature}, assuming that the binarity properties are similar for these populations and a similar spatial resolutions of the data used in the different studies. The conversion of the 25 Ori system IMF to the single-star IMF is beyond the scope of this study.

\subsubsection{Estimation of Missed Members}
\label{sec:missed}
As explained in previous sections, in our estimation of the system IMF we corrected the possible over-counting of individual stars and/or stellar systems belonging to 25 Ori by considering several sources of contamination in the photometric sample. An additional improvement of our procedure is to estimate possible under-counting of members by estimating the number of 25 Ori individual stars and/or stellar systems that could lie outside the PMS locus defined in the CMD.

We made this estimation through a simple simulation of the expected distribution of the cluster members in the $I_c$ vs $I_c-J$ diagram and computing the fraction of these that falls outside the PMS locus. The simulation was performed as follows, in which we refer as \emph{synthetic members} to those individual stars and/or stellar systems obtained from a realization of the system IMF:

$(i)$ We made a random realization of the 25 Ori system IMF by drawing masses for 1000 synthetic members from a lognormal distribution with $m_c=0.31$ and $\sigma=0.46$. These parameters matches the resulting system IMF that will be discussed in Section \ref{sec:parametrizations}. 

$(ii)$ The $I_c$ and $J$-band absolute magnitudes of each synthetic member were computed by interpolating their masses into the mass-luminosity relation using the 7 Myr isochrones of BT-Settl and PARSEC-COLIBRI, as explained in Section \ref{sec:mass-luminosity}.

$(iii)$ The absolute magnitudes were converted into apparent magnitudes by adding the distance moduli and the corresponding extinctions. The distances and visual extinctions were generated for each synthetic member by creating random realizations considering the inversion of the cumulative distributions of the distances from the Gaia DR2 parallaxes and visual extinctions from spectroscopically confirmed members of 25 Ori (see Figure \ref{fig:cum_dist}). Visual extinctions were converted into extinctions in $I_c$ and $J$ bands through the \cite{Rieke-Lebofsky1985} extinction law with $R_V$=3.02.

$(iv)$ We randomly labelled $25$ per cent of the synthetic members as photometrically variables in both $I_c$ and $J$ bands. To each of the variables we assigned a variation, $\Delta I_c$, drawn at random from a normal distribution with zero mean and standard deviation, $\sigma _{I_c}$, equal to 0.3. The fraction of variables as well as $\sigma _{I_c}$ were obtained by matching the catalogue of member candidates with the CVSO, which includes stars and BDs with K and M spectral types. A total of 840 candidates ($\sim$50 per cent of the candidate sample) fainter than $I_c=13$ mag (saturation of the CVSO) have a counterpart in the CVSO and we considered as variable the 220 candidates having a probability $>99$ per cent of being variables in the $I_c$-band. The $J$-band variation was computed by multiplying the $\Delta I_c$ by the ratio between the amplitude variations in the $I_c$ and $J$-bands from \cite{Scholz2009}. Both variations were added to the corresponding apparent magnitudes computed in $(iii)$.

$(v)$ We assumed no IR excesses in the $J$-band because at the 25 Ori age they are observed at larger wavelengths, as explained in Section \ref{sec:excesses}.

$(vi)$ Finally, we simulated the photometric uncertainties in the $I_c$ and $J$-bands by adding to the corresponding apparent magnitude a random error based on an estimation of the photometric errors present in our data. Such estimations were obtained through the fit we did to the mean errors as a function of the mean magnitudes and a fit of the standard deviation of errors as a function of the mean magnitude. Then, for each source, the final apparent magnitude is computed by extracting a magnitude from a normal distribution which is centred at the mean apparent magnitude resulting from $(v)$ with a standard deviation equal to the standard deviation of errors that corresponds to such mean apparent magnitude.

\begin{figure}
	\includegraphics[width=0.47\textwidth]{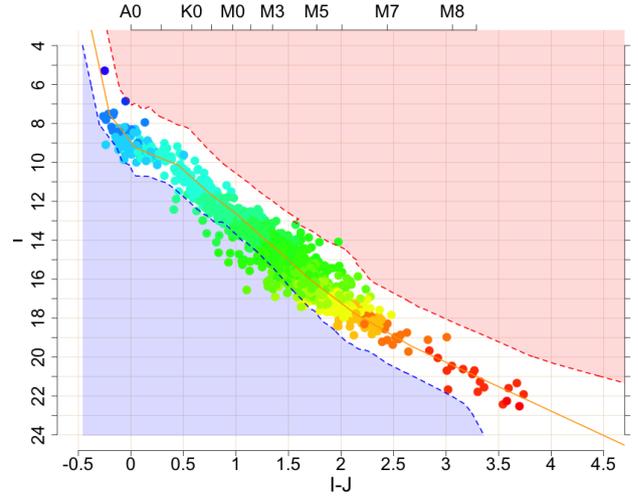}
	\caption{Simulated $I_c$ vs $I_c-J$ diagram for the estimation of the number of members missed by our candidate selection procedure. Dashed lines indicate the PMS locus and solid line the empiric isochrone. The coloured scale indicates the mass of the synthetic members.
	\label{fig:syntheticIvsIJ}}
\end{figure}

We generated 1000 random realizations of the cluster and obtained that a mean fraction of $\sim3.1$ per cent of the synthetic members fall outside the PMS locus, with most of them ($\sim3.0$ per cent) lying on the left side and mainly having $I_c$ magnitudes between 13 and 17 mag. This preferential loss of members does not represent an issue in our system IMF determination because corresponds to changes contained within the uncertainties. We found that this fraction and distribution are consistent with the fraction of confirmed members lying outside the PMS locus shown in Figure \ref{fig:CMD}. In Figure \ref{fig:syntheticIvsIJ} we show the result of a characteristic simulation.

Through the variation of the input parameters within values representative of 25 Ori, we found that the main effects that can move synthetic members outside the PMS locus is the photometric variability. As expected, within reasonable values, the system IMF parameters $m_c$ and $\sigma$ do not affect the number of synthetic members falling outside the PMS locus, so our estimation of the under-counting is not affected by the assumed system IMF.

\subsection{Resulting Sample of Member Candidates}
\label{sec:sample}
The resultant sample selected from the PMS locus in the CMD and after removing potential extragalactic contaminants in the colour-colour diagram has 1687 photometric member candidates with $I_c$ magnitudes between 5.08 and 23.3 mag ($0.011-13.1\ M_\odot$) and covering an area of 1.1$^\circ$ radius in 25 Ori. The completeness of this sample is at $I_c=22.5$ mag ($12\ M_{Jup}$) and the brightest sources in 25 Ori are also included. For a statistical removal of the field star and extragalactic contaminants in this sample, when constructing the LF and system IMF in Sections \ref{sec:LF} and \ref{sec:sys-imf}, respectively, we used the control field and, as a comparison for the galactic contamination, the BGM. The contamination present in our sample depends of the brightness range but it can be roughly characterized into three ranges. The extragalactic contamination starts to be significant for $I_c$ magnitudes fainter than $\sim$17 mag. For the bright $I_c$ range, between $\sim$9 to 13 mag, there is a high level of contamination by giant and subgiant stars (reason why we applied distance and proper motion criteria to filter the sample). In the brightness range between these kind of contaminants, the PMS population is clearly distinguished from the old dwarf stars and the contamination decreases. We estimated, using the control field and/or the BGM, a contamination of $\sim20$ per cent in our sample in the range between 13 and 17 mag. Actually, in this brightness range is where most of the 25 Ori members has been spectroscopically confirmed, as shown in Figure \ref{fig:CMD}.

With our sample we confirmed the low stellar density in 25 Ori. We obtained values between 8.6 and 4.8 stars pc$^{-3}$ for areas with radii between 0.5 and 1.0, while the substellar density ranges from 1.3 to 0.7 BDs {pc$^{-3}$} for the same areas, considering the 25 Ori distance estimated in this study and assuming a spherical group. This stellar density values are roughly consistent with \citet{Briceno2007,Downes2014,Briceno2019}.

We compared our candidate sample with the candidate selection done by \citet{Downes2014} using a similar procedure and the CDSO and VISTA catalogues. Their sample includes candidates with masses in a smaller range ($0.02\le M/M_\odot\le0.80$) but covering a larger area (about 3x3 deg$^2$ around 25 Ori). If we consider the same area as in the present work, there are about 750 candidates in their selection. Our sample contains 924 member candidates in the same mass range and includes 91 per cent of their candidates. From the remaining 9 per cent not included in our sample and with $I_c\ge17$ mag (brightness limit from which we used the DECam photometry), about 85 per cent of them lie outside the DECam detectors, making it imposible to recover those sources in our selection. Thus, where we have full spatial coverage, we recover more than 97 per cent of the member candidates by \citet{Downes2014} and, additionally, we reported 242 new candidates in the same mass range covered by their study. We estimated that the contamination in our candidate sample, in the $I_c$ brightness range between 13 to 20 mag, is about 30 per cent, which is somewhat higher than the 20 per cent estimated and spectroscopically proven by them in their sample. This difference is due, mainly, because our PMS locus is somewhat wider.

\section{Results and Discussion}
\label{sec:results}

\subsection{Luminosity Function}
\label{sec:LF}
In order to construct the LF we calculated the absolute magnitudes of the member candidates and contaminants considering they are real members of 25 Ori. This consideration allow us to analyse properties of the candidate sample as a whole, such as the LF and the system IMF after correcting the contamination effect.

The absolute magnitudes were obtained using our joined $I_c$-band catalogue and, as only 18 per cent of the candidates (those spectroscopically confirmed as members) have visual extinctions from previous studies and 86 per cent of the sample has Gaia DR2 parallaxes with errors of $\le 20$ per cent, we assigned distance and visual extinction values to the whole sample as follow: From a list of 334 spectroscopically confirmed members of 25 Ori \citep{Briceno2005,Briceno2007,Downes2014,Downes2015,Suarez2017,Briceno2019}, we constructed the normalized cumulative distributions of their distances and reported visual extinctions. Then, we used the inversion of these observed distributions to create random realizations to assign values of these parameters to each member candidate, even those already having parallaxes with errors of $\le 20$ per cent or visual extinctions from previous spectroscopic studies in order to have a sample with all values consistent with those of the 25 Ori members. A detailed explanation of this procedure is found in Appendix \ref{sec_app:distance_extinction}. With these distances and extinctions, together with the $I_c$ photometry, we computed the corresponding absolute magnitudes, $M_{I_c}$, for all the member candidates. We made $10^4$ repetitions of this experiment in order to obtain a robust simulation, which produced $10^4$ artificial distributions in the $M_{I_c}$ range from -2.8 to 15.4 mag.

In a similar way we obtained $10^4$ $M_{I_c}$ magnitudes for each candidate in the filtered sample. The resultant $M_{I_c}$ range of this subset is between -2.8 and 9.6 mag, assuming the distance and extinction of 25 Ori, and includes the region mostly affected by giant and subgiant stars.

For the contaminants from the control field and BGM, we estimated their fiducial $M_{I_c}$ magnitudes following the same procedure we used for the member candidates. This way, we can estimate the contamination in the $M_{I_c}$ distribution of the member candidates to then obtain the LF.

Using the simulation just described, we constructed the $10^4$ $M_{I_c}$ distributions of the member candidate and contaminant samples. To correct each distribution by the DECam spatial coverage factor explained in Section \ref{sec:spatial}, we first made the $M_{I_c}$ distributions of the sources from the DECam catalogue and applied them the correction. Then, we added to these distributions those from the rest of the data.

With the $10^4$ $M_{I_c}$ distributions of the member candidate sample we defined the distribution using the mean values and assigning uncertainties of 1-$\sigma$. The errors for the more massive bins, which do not have more than two candidates, are very small because these sources have similar $M_{I_c}$ values for all the repetitions. For these bins we replaced the uncertaintites by the Poisson errors. In a similar way, we defined the distribution of mean values for the candidates in the filtered sample and for the contaminants. The resultant distributions of the contaminants from the control field and the BGM are consistent, within the uncertainties, for $M_{I_c}$ magnitudes brighter than $\sim 9$ mag, even where the giant and subgiant stars lie, which indicates that the contamination in our sample in this range is due mainly to field stars. For fainter sources, a significant discrepancy arises between both samples of contaminants, which increases with the magnitude, suggesting the presence of extragalactic sources. We decided to work with the contaminants estimated from the control field because this also allow us to remove these extragalactic sources.

To the $M_{I_c}$ distribution of the member candidates we subtracted the distribution of the contaminants from the control field and by adding the errors in quadrature. The resultant distribution is very consistent with that from the filtered sample, excluding the $M_{I_c}$ interval ($\sim1-5$ mag) that corresponds to the region presenting a high degree of contamination by giant and subgiant stars. This indicates that the adopted Gaia DR2 thresholds are well-suited. Thus, to obtain the LF of 25 Ori, in the distribution of the member candidates minus the contaminants from the control field we replaced the range $M_{I_c}>5$ mag by the distribution of the candidates in the filtered sample.

In the procedure described above we worked with histograms (with bins of 1 mag) to construct the 25 Ori LF. Additionally, we built the continuous LF of 25 Ori using a kernel density estimate (KDE) and the $M_{I_c}$ magnitudes computed in this section. We obtained the KDEs of the member candidates and contaminants using a bandwidth of 0.4 mag and an Epanechnikov kernel \citep{Silverman1986}. Similarly than for the case of histograms, we first obtained the KDE of the member candidates from the DECam catalogue to apply the correction by the DECam spatial coverage. This KDE was normalized to the histogram of the member candidates at $M_{I_c}=10.5$ mag. Then, we obtained the KDE of the rest of candidates normalized to the same histogram at $M_{I_c}=7.5$ mag. These normalizations allow us to compare the KDEs directly with the histograms. Both KDEs were joined to obtain the KDE of the member candidates for each of the $10^4$ repetitions of the experiment to assign the $M_{I_c}$ magnitudes. Finally, we defined the KDE of the member candidates with the mean values over all the repetitions and assigned uncertainties of 1-$\sigma$. Similarly, we obtained the KDEs of the contaminants from the control field and of the candidates in the filtered sample. The final continuous LF of 25 Ori was obtained subtracting from the KDE of the member candidates the KDE of the contaminants and replacing the KDE of the filtered sample of candidates for $M_{I_c}>5$ mag.

\begin{figure*}
	\centering
	\includegraphics[width=0.95\textwidth]{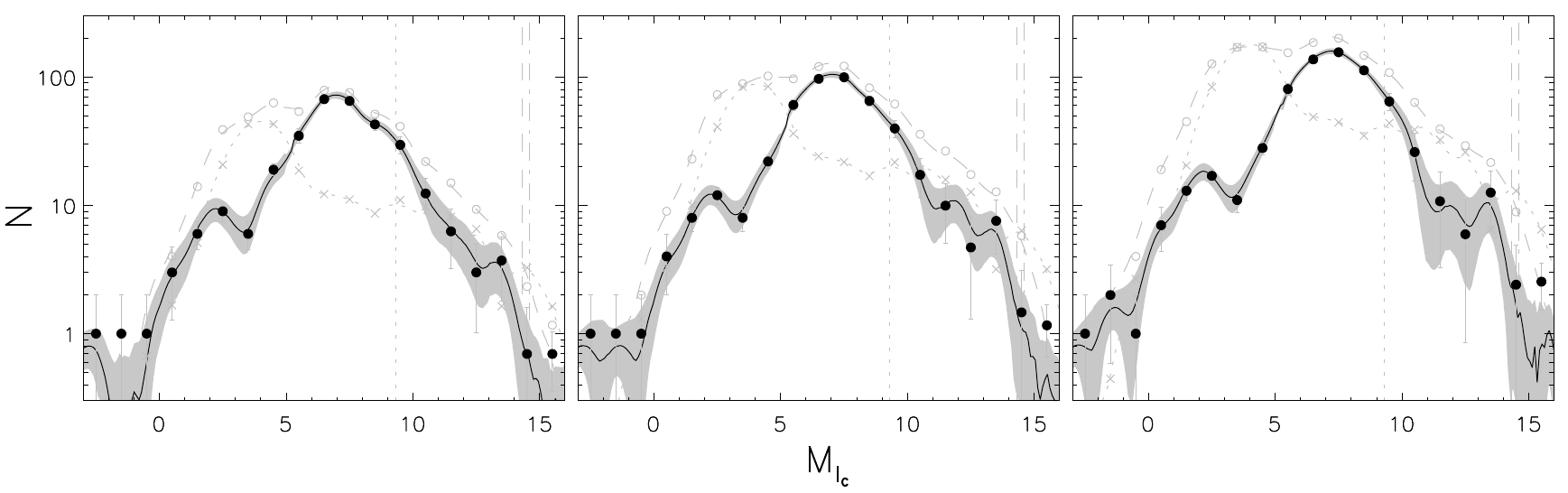}
	\caption{LFs of 25 Ori (black points and black solid curves) after correcting by the galactic and extragalactic contamination (grey crosses and dotted curves) in our member candidate sample (grey open circles and dashed curves). The panels from left to right correspond to the 25 Ori area by \citet[0.5$^\circ$ radius; ][]{Downes2014}, \citet[0.7$^\circ$ radius; ][]{Briceno2019} and \citet[1.0$^\circ$ radius; ][]{Briceno2005,Briceno2007}. The vertical lines, from left to right, indicate the substellar limit (H burning limit), the BD-planetary object limit (D burning limit) and the completeness limit of our DECam observations.
	\label{fig:LF}}
\end{figure*}

We constructed continuous and discrete LFs for different {areas with radius between $0.5$ and $1.1^\circ$}. In Figure \ref{fig:LF} we show the LFs for the 25 Ori estimated areas. These LFs have very similar morphologies, within the uncertainties, regardless the considered area. Also, the KDEs are very consistent with the histograms, specially where we have more than two counts ($M_{I_c}\sim0-14$ mag).

\subsection{System IMF}
\label{sec:IMF}
The main purpose of this study is to determine the system IMF of 25 Ori. Therefore, we need to estimate through a mass-luminosity relationship, the corresponding masses for our member candidates and contaminants under the consideration that both are true members of 25 Ori.

\subsubsection{Mass-Luminosity Relationship}
\label{sec:mass-luminosity}
At the age of 25 Ori \citep[7-10 Myr; ][]{Briceno2005,Briceno2007,Downes2014,Briceno2019}, stars with masses between $\sim 2$ and $\sim 15\ M_\odot$ should be already in the MS, while less massive objects are still in the PMS and more massive stars are in post-MS stages \citep[][]{Prialnik2000}. The most massive star in 25 Ori is the star with the same name, classified as a peculiar B1V star with broad lines \citep{Houk1999}, which roughly corresponds to $\sim 10\ M_\odot$ using the \citet{Schmidt-Kaler1982} empirical mass-luminosity relationship. Therefore, we do not expect in our candidate sample members of 25 Ori being in post-MS but we do expect PMS and MS members. We estimated that $\sim7$ per cent of our candidates have masses larger than 2 $M_\odot$, considering the system IMF by \citet{Downes2014}.

\begin{figure}
	\includegraphics[width=0.47\textwidth]{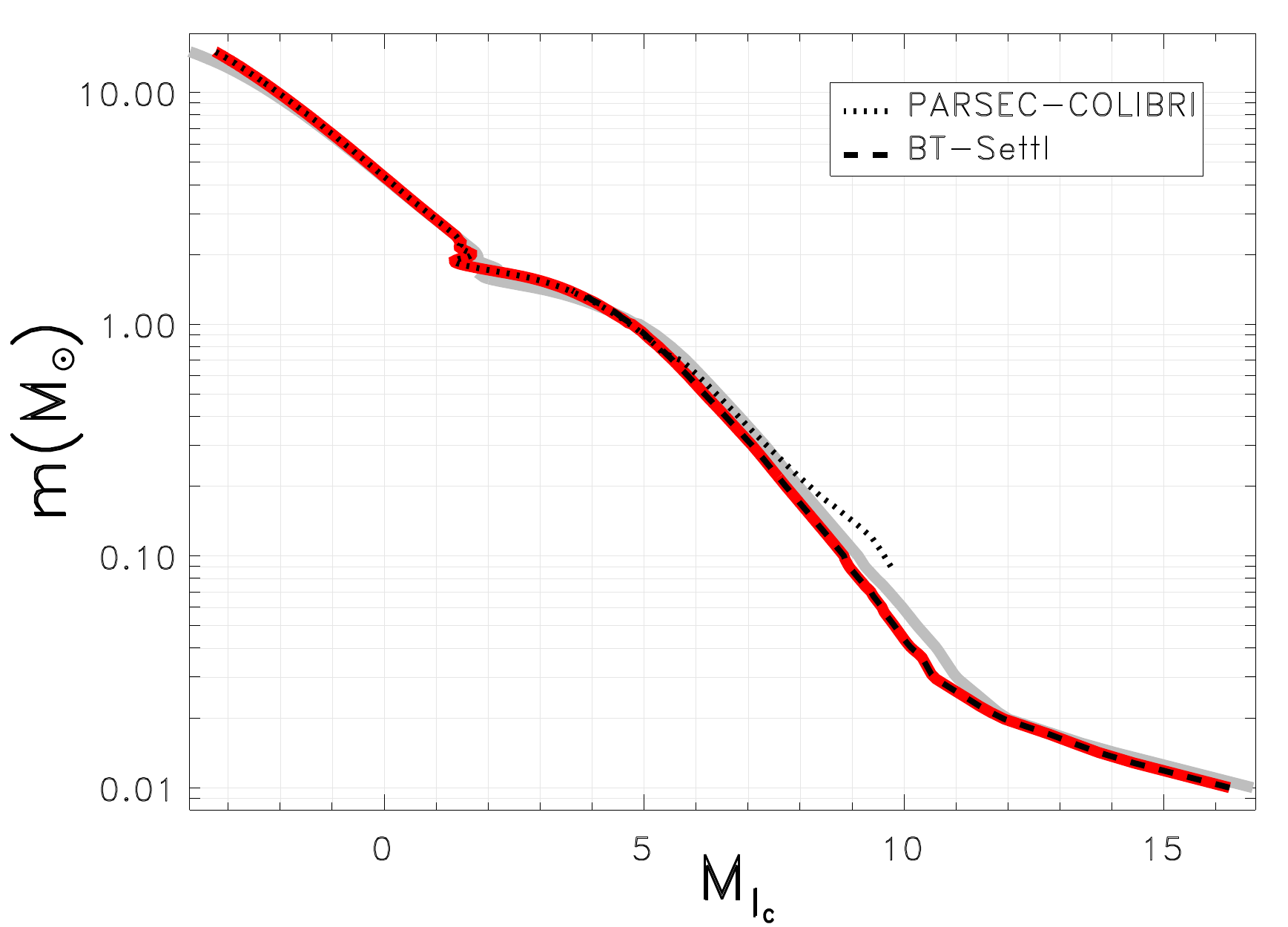}
	\caption{Mass-luminosity relation used to estimate the masses of the member candidates and contaminants (red solid curve). This relation is a combination of the 7 Myr isochrones of BT-Settl and PARSEC-COLIBRI, which are indicated by the dashed curve and the dotted curve, respectively. As a reference, the mass-luminosity relation considering instead the 10 Myr isochrones is represented by the grey solid curve, which is mostly contained into the thickness of the mass-luminosity relation for 7 Myr.}
	\label{fig:mass-L}
\end{figure}

In order to cover the large $M_{I_c}$ range in our candidate sample (from -2.8 to 15.4 mag), we worked with two sets of mass-luminosity relationships for PMS and MS stellar models at the age of 25 Ori. We considered the 7 Myr isochrones of PARSEC-COLIBRI for masses higher than $1.0\ M_\odot$ and of BT-Settl for lower masses. These isochrones were obtained assuming solar metallicity \citep{Biazzo2011b}. In Figure \ref{fig:mass-L} we show the resulting mass-luminosity relation from high-mass stars to very low-mass objects (from 0.01 to 15 $M_\odot$). We stress the soft transition between both isochrones at the selected cutoff (1 $M_\odot$). 

\subsubsection{System IMF Determination}
\label{sec:sys-imf}
By interpolation of the $M_{I_c}$ magnitudes into the mass-luminosity relationship explained in the previous section, we estimated the masses that correspond to each member candidate as well as to each contaminant considering they are members of 25 Ori. Thus, we obtained $10^4$ masses for each source. The resulting mass range covered by the member candidates is between $0.011$ and 13.1 $M_\odot$. In Table \ref{tab:candidates} we list this mass range along with those for the samples of contaminants.

With these masses we constructed the mass distributions of the member candidates and contaminants. Similarly than for the $M_{I_c}$ distributions, we corrected the distributions by the spatial completeness of DECam and then we defined the mass distribution using the mean values and assigning errors of {1-$\sigma$}. For the massive bins, which do not have more than two sources, we replaced the uncertainties by the Poisson errors. From the mass distribution of the member candidates we subtracted that of the control field contaminants adding the errors in quadrature. The resultant distribution is very consistent with that of the candidates in the filtered sample, avoiding the region ($\sim0.8-3\ M_\odot$) with a high degree of contamination by giant and subgiant stars. Thus, we obtained the system IMF of 25 Ori by replacing in the mass distribution of the member candidates minus the contaminants from the control field the range $m>0.8\ M_\odot$ by the distribution of the filtered sample. The derived 25 Ori system IMF is complete from 0.012 to 13.1 $M_\odot$ (corresponding to the 25 Ori star).

Additionally to the discrete determination of the system IMF of 25 Ori, we built the continuous system IMF using a KDE and the masses determined in this section. We assumed a bandwidth of 0.1 dex (in logarithmic scale of mass) and an Epanechnikov kernel \citep{Silverman1986} to obtain the KDEs of the member candidates and contaminants. These KDEs were obtained similarly than for the continuous LFs, normalizing to the mass histograms at $0.03$ and $0.3\ M_\odot$ when working with the sources (candidates or contaminants, depending the KDE to be derived) from the DECam catalogue or from the rest of catalogues, respectively. The final continuous system IMF of 25 Ori was obtained subtracting from the KDE of the candidates the KDE of the contaminants and replacing the KDE of the filtered sample of candidates for $m>0.8\ M_\odot$.

\begin{figure*}
	\centering
	\includegraphics[width=0.95\textwidth]{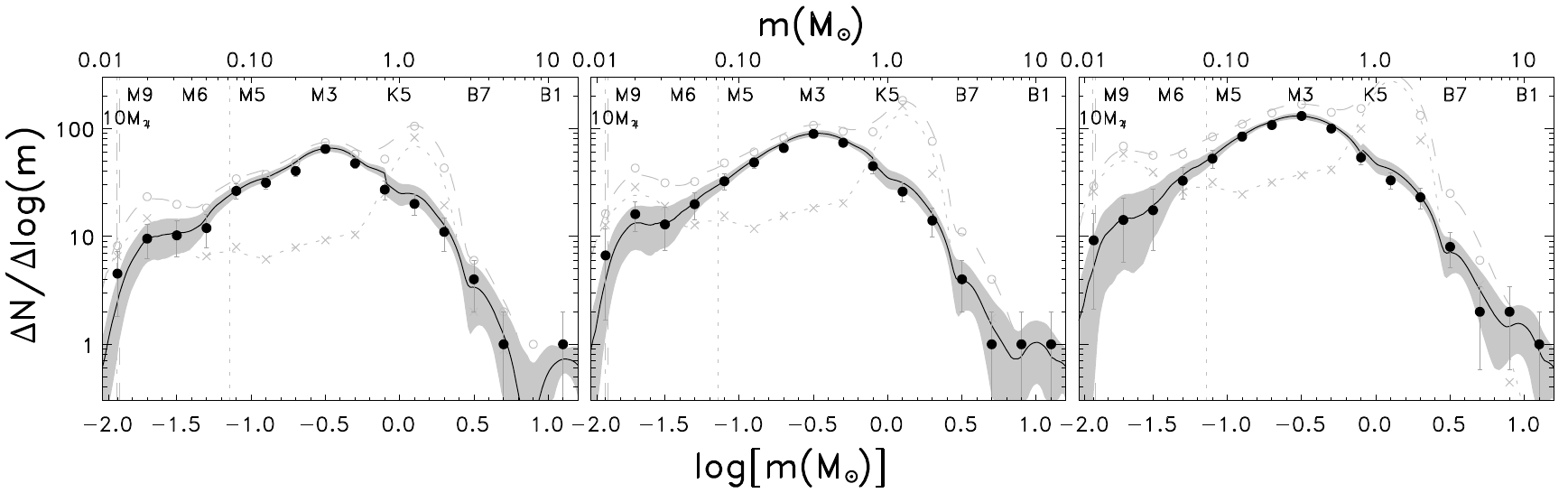}
	\caption{System IMFs of 25 Ori (black points and black solid curves) after correcting by the galactic and extragalactic contamination (grey crosses and dotted curves) in our member candidate sample (grey open circles and dashed curves). The panels, from left to right, correspond to the 25 Ori areas by \citet[0.5$^\circ$ radius; ][]{Downes2014}, \citet[0.7$^\circ$ radius; ][]{Briceno2019} and \citet[1.0$^\circ$ radius; ][]{Briceno2005,Briceno2007}. The vertical lines are the same as in Figure \ref{fig:LF}. The spectral type scale is a combination of the \citet{Pecaut2013} relation and the mass-luminosity relation explained in Section \ref{sec:mass-luminosity}. The size of the bin for the discrete distributions is 0.2 dex.
	\label{fig:imf}}
\end{figure*}

In Figure \ref{fig:imf} we show the system IMFs for the 25 Ori estimated areas. The least massive bin (at log $m=-1.9$) is partially affected by the completeness of our DECam data in the magnitude range between about 21 and 24 mag (DECam completeness at $I_c=22.5$ mag). Then, we corrected the counts in this magnitude range by a factor of $\approx2.5$, which results from the ratio between the expected number of sources (from extrapolation of the linear behaviour of the $I_c$ magnitude distribution in logarithmic scale of the DECam data; see Section \ref{sec:sensitivity}) and those observed with DECam in that magnitude range. We did not correct the least massive interval of the KDE by the DECam incompleteness due to the edges of a KDE are influenced by a boundary effect that occurs in nonparametric curve estimation problems \citep{Silverman1986}. This is not an issue for us because the parametrizations of the system IMF are done considering the discrete determinations. However, we point out the consistency of the histogram and KDE determinations of the system IMF of 25 Ori for the different areas.

\subsubsection{Parametrizations}
\label{sec:parametrizations}
We described the derived system IMF of 25 Ori using the following parametrizations:

$i)$ A two-segment power-law distribution in the form:

\begin{equation}
	\xi(\log m)\propto m^{-\Gamma_i}
\end{equation}

where $\Gamma_1$ and $\Gamma_2$ are the slopes for masses $m<0.40\ M_\odot$ and $m\ge0.4\ M_\odot$, respectively. Such parametrization is inspired by that of the Galactic-field IMF proposed by \citet{Kroupa2001b,Kroupa2002} and by the dual power-law distribution of \citet{Hoffmann2018}, but with a different break mass because these parametrizations are for the single-star IMF.

$ii)$ A lognormal distribution for masses $m\le1M_\odot$, according to \citet{Chabrier2003a,Chabrier2003b}:

\begin{equation}
\xi(\log m)\propto e^{-\frac{(\log m-\log m_c)^2}{2\sigma^2}}
\end{equation}

where $m_c$ is the characteristic mass and $\sigma$ the standard deviation. If we consider the lognormal fit up to 13.1 $M_\odot$, the resultant parameters are in agreement, within the errors, with those when the fit is done for masses $m\le1\ M_\odot$.

$iii)$ A tapered power-law function over the whole mass range of the system IMF ($0.012-13.1\ M_\odot$):

\begin{equation}
\xi(\log m)\propto m^{-\Gamma} \Big[1-e^{-(m/m_p)^\beta}\Big]
\end{equation}

where $m_p$ is the peak mass, $\Gamma$ the power law index and $\beta$ the tapering exponent. This function, introduced by \citet{DeMarchi2005}, has a power law behaviour for high masses and an exponential truncation for lower masses.

\begin{figure*}
	\centering
	\includegraphics[width=0.95\textwidth]{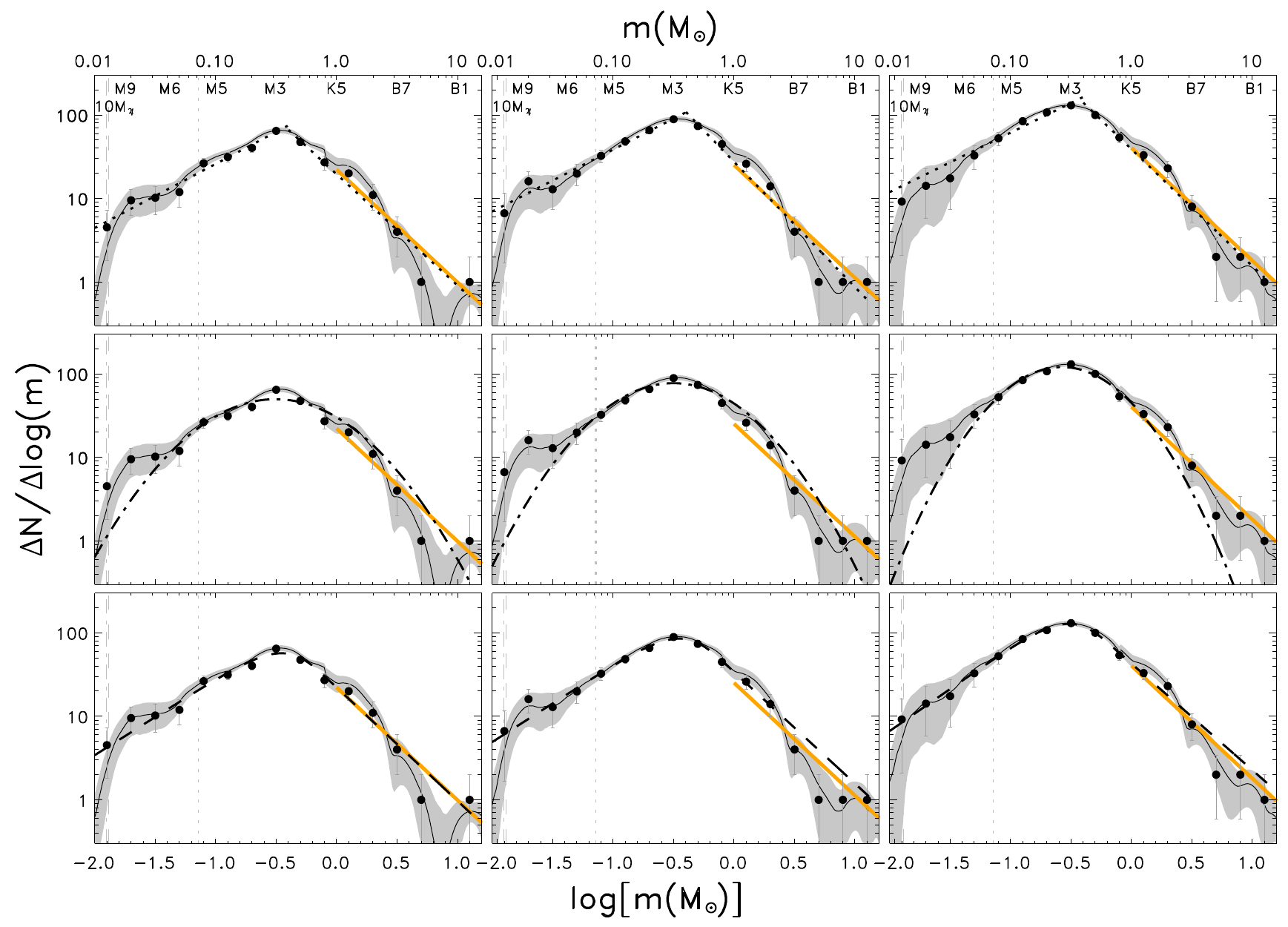}
	\caption{Parameterizations fitted to the 25 Ori system IMFs. Left, central and right panels are the system IMFs considering the areas by \citet[0.5$^\circ$ radius; ][]{Downes2014}, \citet[0.7$^\circ$ radius; ][]{Briceno2019} and \citet[1.0$^\circ$ radius; ][]{Briceno2005,Briceno2007}, respectively. Top, middle and bottom panels show the two-segment power-law, lognormal and tapered power-law functions, respectively, fitted to the system IMFs. As a reference, the orange line shows the \citet{Salpeter1955} slope ($\Gamma=1.35$). The rest of the symbols and lines are the same as in Figure \ref{fig:imf}.
	\label{fig:imf_par}}
\end{figure*}

The fits were done in the discrete determination of the system IMF, as mentioned in Section \ref{sec:sys-imf}. In Figure \ref{fig:imf_par} we show the parametrizations of the 25 Ori system IMF and in Table \ref{tab:imf} we summarize the parameters with their uncertainties. 

\begin{table*}
\caption{Parameterizations fitted to the 25 Ori system IMF.}
  \small
  \label{tab:imf}
  \begin{threeparttable}
 	\begin{tabular}{@{\extracolsep{2pt}}cccccccc@{}}
    \toprule
		Area           & \multicolumn{2}{c}{Lognormal} & \multicolumn{2}{c}{Two-Segment Power Law}           & \multicolumn{3}{c}{{Tapered} Power-Law} \\
   \cline{2-3}
   \cline{4-5}
   \cline{6-8}
        radius	    & $m_c$        & $\sigma$       & $\Gamma_1$ ($m<0.4\ M_\odot$) & $\Gamma_2$ ($m\ge0.4\ M_\odot$) & $\Gamma$      & $m_p$         & $\beta$      \\
        ($^\circ$)  & $(M_\odot)$  &                &                               &                &                & $(M_\odot)$   &             \\
    \midrule
		0.5$^a$       & 0.31$\pm$0.06 & 0.51$\pm$0.08 & -0.77$\pm$0.06 & 1.33$\pm$0.12 & 1.36$\pm$0.39 & 0.36$\pm$0.07 & 2.27$\pm$0.33 \\
		0.7$^b$       & 0.32$\pm$0.04 & 0.47$\pm$0.06 & -0.74$\pm$0.04 & 1.50$\pm$0.11 & 1.34$\pm$0.14 & 0.36$\pm$0.03 & 2.26$\pm$0.11 \\
		1.0$^c$       & 0.27$\pm$0.02 & 0.41$\pm$0.03 & -0.71$\pm$0.07 & 1.40$\pm$0.09 & 1.28$\pm$0.07 & 0.30$\pm$0.02 & 2.28$\pm$0.07 \\
    \bottomrule
 	\end{tabular}
  \begin{tablenotes}[para,flushleft]
	$^a$By \citet{Downes2014}.\\
	$^b$By \citet{Briceno2005,Briceno2007}.\\
	$^c$By \citet{Briceno2019}.
  \end{tablenotes}
 \end{threeparttable}
\end{table*}

\subsubsection{Comparison of the 25 Ori system IMF with Other Studies}
\label{sec:imf_comparison}
Before comparing the system IMF reported here with that in other regions, we considered the 25 Ori system IMF obtained by \citet{Downes2014}. They found that the system IMF in their entire survey (3x3 deg$^2$ around 25 Ori) is well described by either two power laws with slopes $\Gamma_a=-2.73\pm0.31$ and $\Gamma_b=-0.32\pm0.41$ for the mass ranges $0.02\le m/M_\odot\le0.08$ and $0.08\le m/M_\odot\le0.5$, respectively, or a lognormal function with parameters $m_c=0.21\pm0.02$ and $\sigma=0.36\pm0.03$ over the whole studied mass range. Additionally, for the system IMF of the overdensity (0.5$^\circ$ radius), they obtained $\Gamma_a=-2.97\pm0.02$ and $\Gamma_b=-0.63\pm0.04$, and $m_c=0.22\pm0.02$ and $\sigma=0.42\pm0.05$ in the corresponding mass ranges. Those $m_c$ and $\sigma$ values are slightly lower and the slope for substellar masses is quite steeper than those reported here. We mainly attribute these differences between both system IMFs to differences in the corresponding samples and also in the procedures used in both works. Here, we considered the mass range $0.01<M/M_\odot<13$ against $0.03<M/M_\odot<0.8$ from \citet{Downes2014}, in which some level of incompleteness was expected in the less and more massive system IMF bins. Particularly, we estimated that $\sim$10 {per cent} of our member candidates with $I_c$ magnitudes between 17 and 19 mag are unresolved sources in the CDSO catalogue used by \citet{Downes2014}, which could result in that fraction of missed candidates in their selection in this brightness range due to the spatial resolution differences between the DECam and CDSO catalogues, as shown in Table \ref{tab:catalogues}. Additionally, both system IMF estimations followed different procedures: \citet{Downes2014} interpolates masses simultaneously from $T_{eff}$ and $L_{bol}$ in the H-R diagram while here we obtained the system IMF using the mass-luminosity relationship explained in Section \ref{sec:mass-luminosity}. Thus, in this work we present the updated version of the system IMF of 25 Ori over its whole mass range, which allows us to rule-out the possible low number of BDs suggested by \citet{Downes2014} when comparing with the Galactic-disk IMF from \citet{Chabrier2003b}.

In order to contribute to the understanding of the origin of the IMF and its relation with environmental conditions, we compared the parameters of the two-segment power-law and lognormal functions fitted to the 25 Ori system IMF with those in Table \ref{tab:imf_literature}, mainly because those IMFs cover a wide mass range as that presented here, and with other studies of interest to include as well the tapered power-law parametrizations. In these comparisons we assumed similar binarity properties for the different clusters and similar spatial resolutions of the surveys.

The best fitted lognormal function to the 25 Ori system IMF are roughly consistent, within the uncertainties, with those obtained in the clusters mentioned in Table \ref{tab:imf_literature}. The values of $m_c$ range from 0.25 to 0.36 $M_\odot$, with the most widely varying values in the oldest associations (Blanco 1 and Pleiades). $\sigma$ takes values between 0.44 and 0.58 (considering the parametrizations of $\sigma$ Ori for $m<$1 $M_\odot$ and of the ONC with the full sample by \citealt{DaRio2012}). Also, these values are consistent with a set of young clusters in \citet{Bayo2011}. Though we compared the best fitting lognormal function, we point out that this functional form tends to underestimate the number of BDs in 25 Ori. Similar results were reported in $\sigma$ Ori by \citet{PenaRamirez2012} and Upper Sco by \citet{Lodieu2013}, and are predicted by \citet{Hoffmann2018}.

From the power-law fit, the slope we obtained for LMSs and intermediate/high-mass stars is very consistent with the \citet{Salpeter1955} slope ($\Gamma=1.35$) and with the most representative slope for $m\ge1\ M_\odot$ of a large sample of stellar associations in \citet{Bastian2010}, which cover a diversity of physical conditions such as age, metallicity and total mass. However, this slope is slightly steeper than those for most clusters in Table \ref{tab:imf_literature}, although in agreement with the ONC (considering the sample without age threshold by \citealt{DaRio2012}) and with the intermediate-mass slope of Pleiades \citep{Bouy2015}. In order to check the possibility that our slope $\Gamma_2$ may be affected by missed massive members of 25 Ori in our PMS locus selection, we applied the Gaia DR2 criteria to all the brightest sources ($I_c\lesssim10$ or $J\lesssim10$ mag) in our optical catalogue (from UCAC4 and $Hipparcos$) and NIR catalogue (from 2MASS). All the sources satisfying the adopted Gaia DR2 thresholds lie inside our PMS locus, which indicate we are including all massive members of 25 Ori in our candidate selection.

In the case of the slope for the very low-mass and BD regime, the value we obtained for 25 Ori is roughly consistent with most of the regions in Table \ref{tab:imf_literature} and in Table 4 of \citet{Muzic2017}. The slope ranges between -0.3 and -0.8, which restricts the fall (in logarithmic scale of mass) of the number of very LMSs and BDs with mass in a star-forming region with respect to the review of \citet{Bastian2010}. A significant steeper slope is reported in the ONC \citep{DaRio2012}, but a flatter value consistent with the mentioned range is also obtained in that same cluster \citep{Muench2002,Lucas2005}. Also, we observe that the flattest slope reported in the contributions summarized in the mentioned tables is in NGC 1333 when considering masses up to 1 $M_\odot$ \citep{Scholz2013} but, when considering masses $m<0.6\ M_\odot$ the slope is consistent with the range quoted here \citep{Scholz2012}. Thus, when comparing slope fits it is important to take care of the mass range that was considered in each study, specially if the fit extends for masses somewhat higher than the characteristic mass. In the case of 25 Ori, the slope for very LMSs and BDs do not present significant differences if the fit is done in any mass range with masses lower than 0.5 $M_\odot$, but it significantly flattens by roughly 50 per cent if higher masses are considered in the fit.

About the tapered power-law fit to our system IMF, it is roughly consistent with that reported in an extended sample of young clusters (25 Ori not included) by \citet{DeMarchi2010} and \citet{Bastian2010}, which has the parameters $\Gamma=1.1\pm0.2$, $m_p=0.23\pm0.10$ and $\beta=2.4\pm0.4$. The $m_p$ value is slightly higher in our system IMF but the differences are in agreement within the errors.

These comparisons indicate that the 25 Ori system IMF is similar to that of a diversity of stellar clusters, which supports the idea that the shape of the IMF is largely insensitive to environmental properties, as predicted by the models from \citet{Bonnell2006}, \citet{Elmegreen2008} and \citet{Lee-Hennebelle2018b}.

Also, we emphasize that the 25 Ori system IMF is a smooth function across the entire mass range, in the sense that we do not observe any bimodality behaviour as in the ONC \citep{Drass2016}.

\subsection{BD/star ratio}
\label{sec:BD_star_ratio}
An alternative quantity that indicates the relative efficiency to form stellar and substellar objects is, precisely, the ratio between BDs and stars. We worked with the $R_{ss}$ definition by \citet{Briceno2002}, which is the ratio between BDs and stars considering objects with masses between 0.02 and 10 $M_\odot$ and the BD-star limit at 0.08 $M_\odot$. For the 25 Ori areas of 1.0, 0.7 and 0.5$^\circ$ radius, $R_{ss}$ is $0.15\pm0.03$, $0.16\pm0.02$, $0.16\pm0.02$, respectively. Similar $R_{ss}$ values are obtained for other radii between 0.4 and 1.1$^\circ$.

The $R_{ss}$ value representative of 25 Ori is $0.16\pm0.03$ i.e. for each 6 stars in 25 Ori we roughly expect 1 BD. This value is consistent with those found in regions with low stellar density as Blanco 1 \citep{Moraux2007a} and with higher stellar density such as NGC 6611 \citep{Oliveira2009}, and is somewhat lower but consistent within the uncertainties, with those in higher stellar density regions such as the Trapezium \citep{Muench2002,Thies-Kroupa2007}, ONC \citep{Kroupa-Bouvier2003}, IC 348 \citep{Scholz2013}, Chamaeleon-I and Lupus-3 \citep{Muzic2015}. Also, the BD/star ratio we found in 25 Ori is consistent with that on the Galactic plane \citep{Bihain-Scholz2016}. The fact that such widely differing regions show similar ratios of BDs to stars suggests that the environment plays a small role, if any, in the formation of substellar and stellar objects.

We point out that not all the mentioned BD/star ratios were estimated assuming the same mass ranges. In fact, we observed that most of the slightly higher reported values with respect to that obtained here considered masses lower than 1 $M_\odot$ (as suggested by \citealt{Andersen2008}), while we included sources up to 10 $M_\odot$ (as defined by \citealt{Briceno2002}). However, in the case of 25 Ori and considering the \citet{Andersen2008} definition, we obtained a BD/star ratio of $0.17\pm0.03$, which is very consistent with that we obtained using the \citet{Briceno2002} definition.

\subsection{Spatial Distribution}
\label{sec:spatial_distribution}
Taking advantage of the large spatial coverage of our candidate sample, we examined the system IMF for possible variations with the radius. In Table \ref{tab:imf} and Figure \ref{fig:imf_par} we observe that the system IMFs for the different 25 Ori estimated areas are very consistent according to all the parametrizations. Similar results are obtained for other radii between 0.4 and 1.1$^\circ$ (2.5-6.8 pc), which suggests that the stellar and substellar populations of 25 Ori do not have any preferential spatial distribution.

Additionally, the similar $R_{ss}$ values obtained in previous section are an indicative that the substellar and stellar population have similar spatial distribution over the full area of 25 Ori.

We can see in Figure \ref{fig:sky} two stellar groups surrounding 25 Ori whose members could affect the determination of the 25 Ori system IMF. HR 1833 is a prominent overdensity in the spectroscopic study of \citet{Briceno2019} while ASCC 18, detected by \citet{Kharchenko2005}, is not an obvious overdensity in \citet{Briceno2019} and is present very faintly in \citet{Zari2017}. The reported radii of HR 1833 is 0.5$^\circ$ \citep{Briceno2019} and 0.44$^\circ$ for ASCC 18 \citep{Kharchenko2013}. If we consider the 25 Ori area of 0.5$^\circ$ radius \citep{Downes2014}, neither HR 1833 nor ASCC 18 overlap 25 Ori, which allow us an analysis including only member candidates lying inside 25 Ori. Thus, the system IMF for this radius is representative of 25 Ori and do not present significant variations for radius up to 1.1$^\circ$, where somewhat contamination by surrounding groups could be present.

\subsection{Gravitational State of 25 Ori}
As found by \citet{Lada-Lada2003} and predicted by \citet{Bonatto-Bica2011}, most clusters are dissolved before they reach an age of 10 Myr; only less than 10 per cent reach older ages and about 4 per cent survive longer than 100 Myr. 25 Ori is just at this critical point and no conclusive results about its gravitational state have been presented \citep{McGehee2006,Downes2014}. Any cluster, to be gravitationally bound, its escape velocity, $v_{esc}=(2GM/R)^{1/2}$, must be larger than its velocity dispersion \citep{Sherry2004}.

Directly counting in the mass distributions shown in Figure \ref{fig:imf}, we obtained a total mass of $158\pm18$, $223\pm21$ and $324\pm25$ $M_\odot$ contained in 25 Ori inside areas of 0.5, 0.7 and 1.0$^\circ$ radius, respectively. The fraction of these masses contained in BDs is $1.42\pm0.45$, $1.41\pm0.40$ and $1.46\pm0.35$ per cent, respectively. Similar values are obtained for other radius between 0.4 and 1.1$^\circ$, which also indicates, as from the $R_{ss}$ ratio, alike spatial distribution of the substellar and stellar population of 25 Ori.

Considering the total mass of 324 $M_\odot$ inside a radius of 1.0$^\circ$, which corresponds to 6.2 pc at a distance of 356 pc, the resultant $v_{esc}$ is 0.7 km s$^{-1}$. A similar $v_{esc}$ is obtained if considering the total mass inside the 0.7 or 0.5$^\circ$ radii. This $v_{esc}$ is about 3 times smaller than the velocity dispersion of 2 km s$^{-1}$ in 25 Ori \citep{Briceno2007}, which indicates that 25 Ori is an unbound association. We estimated that to be a gravitationally bound cluster, 25 Ori should have about 10 times more mass than that estimated here, which implies an unrealistic number of more than 6000 members, or to have a significantly smaller velocity dispersion.

\section{Summary and Conclusions}
\label{sec:conclusions}
By combining optical and NIR photometry from DECam, CDSO, UCAC4 and $Hipparcos$, and VISTA and 2MASS, respectively, we selected a sample of 1687 photometric member candidates in an area of 1.1$^\circ$ radius in 25 Ori on the basis of their position in colour-magnitude and colour-colour diagrams. This sample covers an $I_c$ range between 5.08 and 23.30 mag, which corresponds to a mass range from $0.011$ to 13.1 $M_\odot$. The completeness of the sample is at 0.012 $M_\odot$, which is just beyond the Deuterium burning limit (0.013 $M_\odot$), and also  includes the most massive stars in 25 Ori. We estimated a contamination of 20 per cent for the LMS candidates, but it increases for the intermediate-mass candidates due to giant and subgiant stars and for BD candidates due to extragalactic sources.

Additionally, we discussed and/or considered, in the context of 25 Ori, the following uncertainties and biases to be taken into account when determining the mass distribution: spatial completeness, photometric sensitivity, IR excesses, chromospheric activity, unresolved binaries and missed members.

With the sample of member candidates we constructed the system IMF of 25 Ori for different areas, which is complete down to 0.012 to 13.1 $M_\odot$ and is one of the few system IMFs over the whole mass range of a stellar cluster (e.g. Collinder 69 by \citealt{Bayo2011} and $\sigma$ Ori by \citealt{PenaRamirez2012}). This system IMF is a smooth function across the entire mass range. We parametrized the resultant system IMF using a two-segment  power-law, a lognormal and a tapered power-law function to compare it with other studies. We observed that a lognormal function well-fitted to the peak of the mass distribution underestimates the BD population of 25 Ori.

The system IMF presented here shows a larger number of BDs than that reported by \citet{Downes2014}. We found this difference can be mainly explained by issues related to the spatial resolution and completeness of the CDSO as well as differences in the procedures for computation of the system IMF. The updated system IMF presented in this work allows us to rule-out the possible low number of BDs suggested by \citet{Downes2014}.

The 25 Ori system IMF does not present significant differences in comparison with other clusters having different physical properties, which suggests that the conversion of gas into stars and BDs has minimum influence by the environmental properties, as predicted by some models \citep[e.g. ][]{Bonnell2006,Elmegreen2008,Lee-Hennebelle2018b}.

We estimated the BD/star ratio of 25 Ori, which has a representative value of $0.16\pm0.03$. This ratio is roughly consistent with those in other regions with different stellar densities which is an indicative that the formation of BDs and stars occurs in a similar way in different environments.

There are no significant variations of the 25 Ori system IMF with radius and the BD/star ratio is similar for different radii between 2.5 to 6.8 pc (0.4-1.1$^\circ$). These results indicate that the substellar and stellar objects do not have any preferential spatial distribution.

Comparing the escape velocity estimated for 25 Ori and its velocity dispersion, we found that 25 Ori is an unbound association. In fact, 25 Ori should have about 10 times more mass or a significantly smaller velocity dispersion to be considered as a gravitationally bound cluster.

The system IMF of 25 Ori we present in this work was constructed with photometric member candidates. To determine the membership of each candidate it is necessary a follow up spectroscopy. Thus, we could determine the distribution of the masses of the confirmed members. This kind of study requires the use of several multi-fiber spectrographs to have full coverage of the brightness range and spatial distribution. In this direction, we have an ongoing spectroscopic survey about 85 per cent complete, which will be part of a future work. 

\section*{Acknowledgments}

We thank the referee for helpful comments and suggestions. We also thank to Amelia Bayo, Jes\'us Hern\'andez and Javier Ballesteros-Paredes for valuable comments on an early version of the manuscript and to Annie C. Robin for support on the application of the Besan\c{c}on Galactic model.

GS acknowledges support from a CONACYT/UNAM fellowship and from the Posgrado en Astrof\'isica graduate program at Instituto de Astronom\'ia, UNAM.
GS, CRZ and JJD acknowledge support from programs UNAM-DGAPA-PAPIIT IN116315 and IN108117, Mexico.
JJD acknowledges support from Secretar\'ia de Relaciones Exteriores del Gobierno de M\'exico and Programa de Movilidad e Intercambios Acad\'emicos de la Comisi\'on Sectorial de Investigaci\'on Cient\'ifica, Universidad de la Rep\'ublica, Uruguay.
MC acknowledge financial support from the Spanish Ministry of Economy and Competitiveness (MINECO) under the grants AYA2015-68012-C2-01, AYA2014-58861-C3-1-P and AYA2017-88007-C3-1-P.
This work is based on observations obtained with VISTA under ESO-program ID 60.A-9285(B).

\par
Based on observations obtained at the Llano del Hato National Astronomical Observatory of Venezuela, operated by the Centro de Investigaciones de Astronom{\'\i}a (CIDA) for the Ministerio del Poder Popular para Educaci\'on Universitaria, Ciencia y Tecnolog{\'\i}a.

\par
This project used data obtained with the Dark Energy Camera (DECam), which was constructed by the Dark Energy Survey (DES) collaborating institutions: Argonne National Lab, University of California Santa Cruz, University of Cambridge, Centro de Investigaciones Energ\'eticas, Medioambientales y Tecnol\'ogicas-Madrid, University of Chicago, University College London, DES-Brazil consortium, University of Edinburgh, ETH-Zurich, University of Illinois at Urbana-Champaign, Institut de Ciencies de l\'Espai, Institut de Fisica d\'Altes Energies, Lawrence Berkeley National Lab, Ludwig-Maximilians Universitat, University of Michigan, National Optical Astronomy Observatory, University of Nottingham, Ohio State University, University of Pennsylvania, University of Portsmouth, SLAC National Lab, Stanford University, University of Sussex, and Texas A\&M University. Funding for DES, including DECam, has been provided by the U.S. Department of Energy, National Science Foundation, Ministry of Education and Science (Spain), Science and Technology Facilities Council (UK), Higher Education Funding Council (England), National Center for Supercomputing Applications, Kavli Institute for Cosmological Physics, Financiadora de Estudos e Projetos, Funda\c{c}\~ao Carlos Chagas Filho de Amparo a Pesquisa, Conselho Nacional de Desenvolvimento Cient\'ifico e Tecnol\'ogico and the Minist\'erio da Ciencia e Tecnologia (Brazil), the German Research Foundation-sponsored cluster of excellence \"Origin and Structure of the Universe\" and the DES collaborating institutions.

\par
We thank the assistance of the personnel, observers, telescope operators and technical staff at CIDA and CTIO who made possible the observations at the J\"urgen Stock telescope at the Venezuela National Astronomical Observatory (OAN) and Blanco Telescope at CTIO.

\par
This publication makes use of data products from the Two Micron All Sky Survey, which is a joint project of the University of Massachusetts and the Infrared Processing and Analysis Center/California Institute of Technology, funded by the National Aeronautics and Space Administration and the National Science Foundation.

\par
This work has made use of data from the European Space Agency (ESA) mission
{\it Gaia} (\url{https://www.cosmos.esa.int/gaia}), processed by the {\it Gaia}
Data Processing and Analysis Consortium (DPAC,
\url{https://www.cosmos.esa.int/web/gaia/dpac/consortium}). Funding for the DPAC
has been provided by national institutions, in particular the institutions
participating in the {\it Gaia} Multilateral Agreement.

\par
Funding for the SDSS and SDSS-II has been provided by the Alfred P. Sloan Foundation,
the Participating Institutions, the National Science Foundation, the U.S. Department
of Energy, the National Aeronautics and Space Administration, the Japanese Monbukagakusho,
the Max Planck Society, and the Higher Education Funding Council for England. The SDSS
Web Site is \url{http://www.sdss.org/}. The SDSS is managed by the Astrophysical Research
Consortium for the Participating Institutions. The Participating Institutions are the
American Museum of Natural History, Astrophysical Institute Potsdam, University of Basel,
University of Cambridge, Case Western Reserve University, University of Chicago,
Drexel University, Fermilab, the Institute for Advanced Study, the Japan Participation
Group, Johns Hopkins University, the Joint Institute for Nuclear Astrophysics, the
Kavli Institute for Particle Astrophysics and Cosmology, the Korean Scientist Group,
the Chinese Academy of Sciences (LAMOST), Los Alamos National Laboratory,
the Max-Planck-Institute for Astronomy (MPIA), the Max-Planck-Institute for
Astrophysics (MPA), New Mexico State University, Ohio State University, University
of Pittsburgh, University of Portsmouth, Princeton University, the United States Naval
Observatory, and the University of Washington.

\par
This work makes extensive use of the following tools: TOPCAT and STILTS available
at http://www.starlink.ac.uk/topcat/ and http://www.starlink.ac.uk/stilts/, R from
the R Development Core Team (2011) available at http://www.R-project.org/ and described
in \emph{R: A language and environment for statistical computing} from R Foundation
for Statistical Computing, Vienna, Austria. ISBN 3-900051-07-0, IRAF which is
distributed by the National Optical Astronomy Observatories, which are operated by
the Association of Universities for Research in Astronomy, Inc., under cooperative
agreement with the National Science Foundation.



\bibliographystyle{mnras}
\bibliography{mybib_Suarez} 

\appendix

\section{Calibration of the DECam Photometry}
\label{sec_app:DECam_calibration}
To calibrate our DECam photometry we first added the zero point of 25.18 mag from the image headers to the instrumental magnitudes. Then, we compared these instrumental magnitudes with the $i$ magnitudes in the DECam system obtained using the $i$ and $z$-band photometry from SDSS according to Transformation \ref{eq:DECam_instrumental}\footnote{\url{http://www.ctio.noao.edu/noao/content/Photometric-Standard-Stars-0\#transformations}}.

\begin{equation} \label{eq:DECam_instrumental}
	i_{DECam}    = i + 0.014 - 0.214*(i-z) - 0.096*(i-z)^2
\end{equation}

where $i_{DECam}$ are in the DECam system and $i$ and $z$ magnitudes are in the SDSS system.

\begin{figure}
	\centering\includegraphics[width=0.40\textwidth]{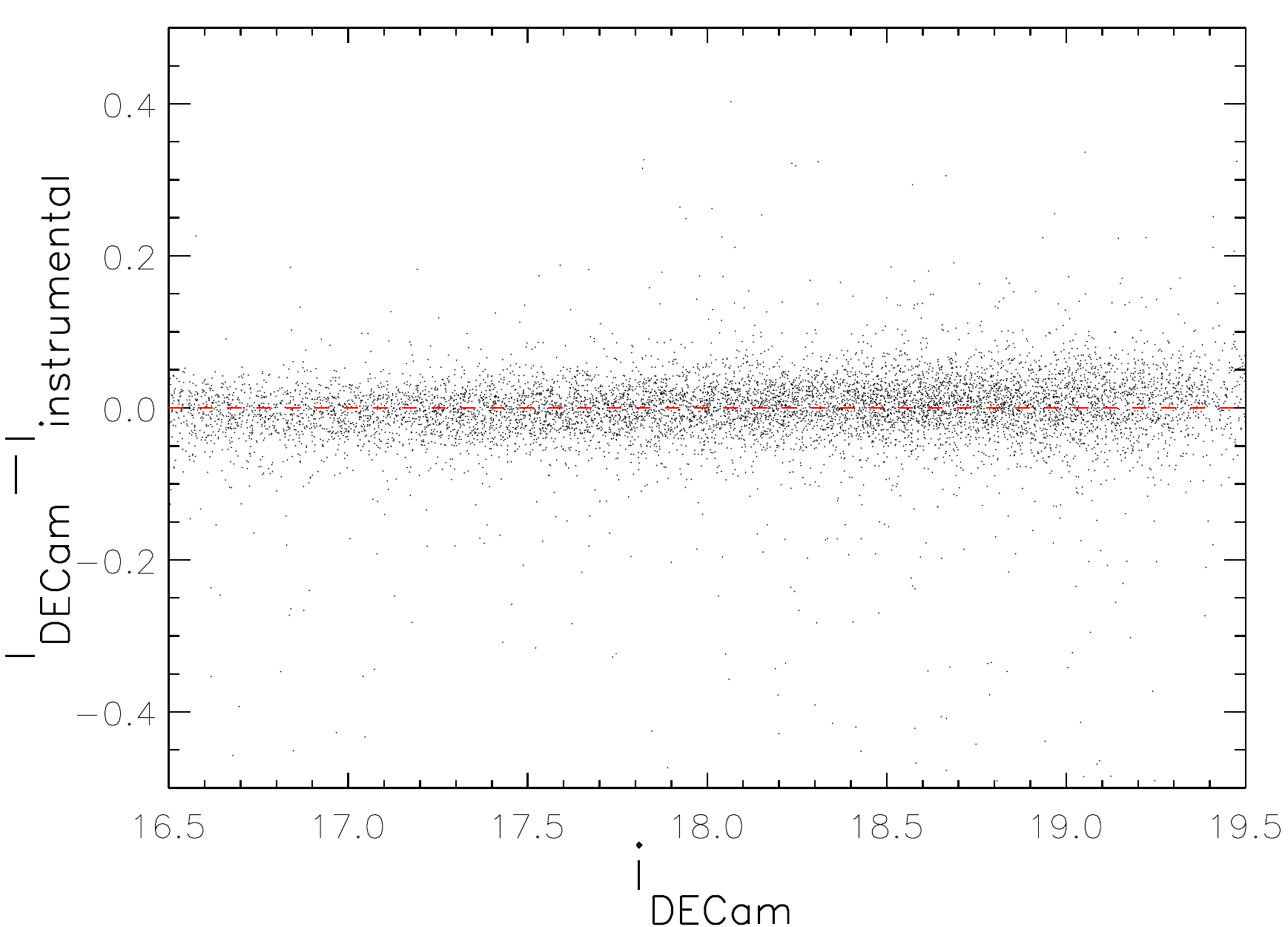}
	\caption{Residual between our calibrated photometry from DECam and the $i$-band photometry in the DECam system obtained using the SDSS catalogue.}
	\label{fig:DECam_instrumental}
\end{figure}

The comparison was done for sources having colours $i-z<0.8$ mag (valid range of the transformation), considering sources not having a high probability of being variable stars according to the CIDA Variability Survey of Orion \citep[][]{Briceno2005,Mateu2012,Briceno2019} and for sources having $i$ and $z$-band photometric errors lesser than 0.05 mag. The mean value of the resultant residuals is 0.637 mag. Thus, we added this value to our DECam photometry to calibrate it. In Figure \ref{fig:DECam_instrumental} we show the residual between our calibrated photometry and that in the DECam system using the SDSS catalogue. The typical residuals are -0.001 mag with a RMS of 0.038 mag.

\section{Transformation of the UCAC4 and DECam Photometry to $I_c$ Magnitudes}
\label{sec_app:photometry_transformation}
We used transformations from \citet{Jordi2006} and empirical relations obtained directly from our data to convert the $i$-band magnitudes from the UCAC4 and DECam catalogues to the $I_c$-band magnitudes.

\subsection{UCAC4 Data}
As the transformations from \citet{Jordi2006} relate the SDSS and Cousins photometric systems, we first checked that the UCAC4 photometry are in the SDSS system.

\begin{figure*}
	\centering
	\begin{tabular}{cc}
		\subfloat{\includegraphics[width=.40\linewidth]{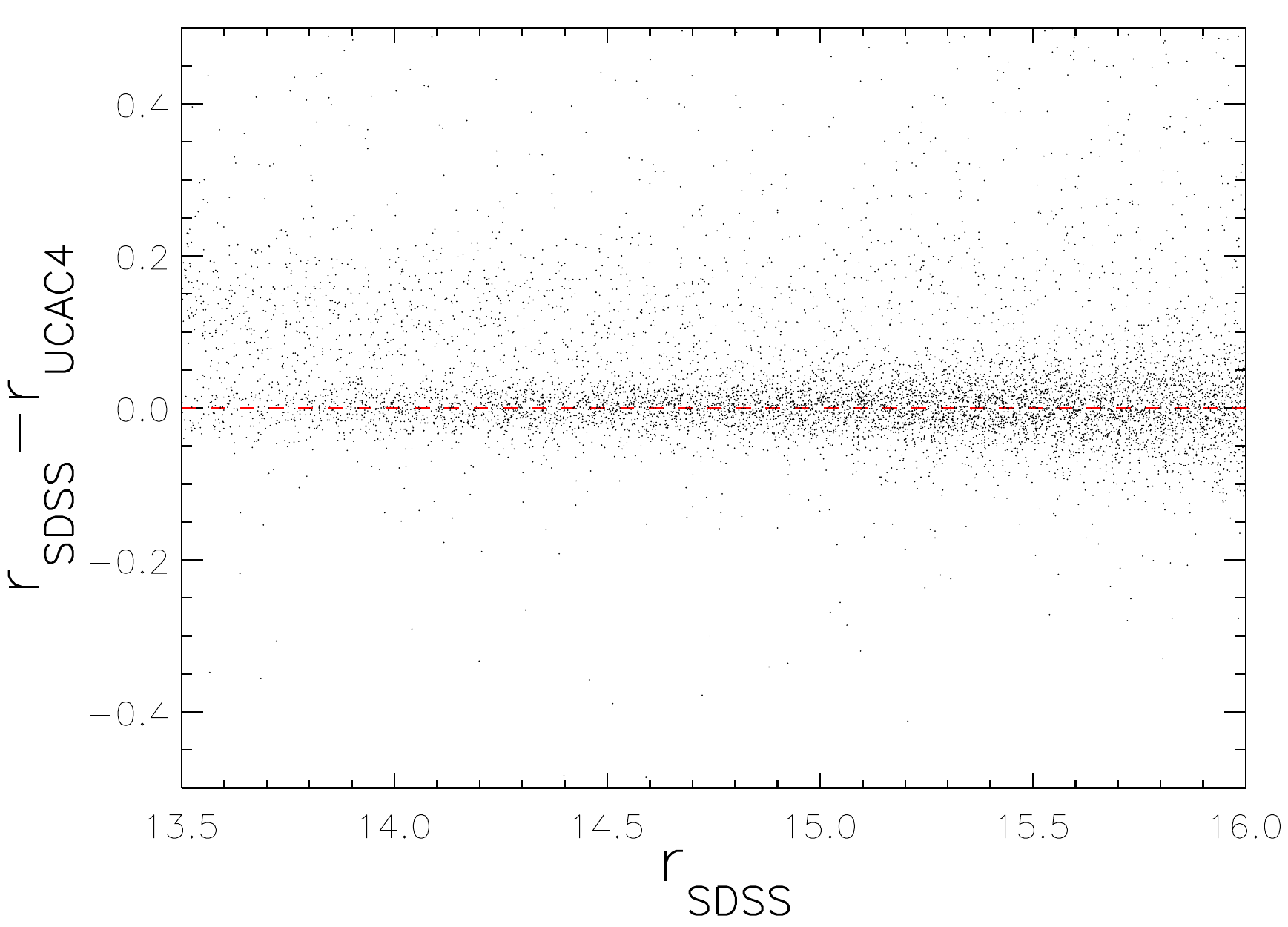}} & 
		\subfloat{\includegraphics[width=.40\linewidth]{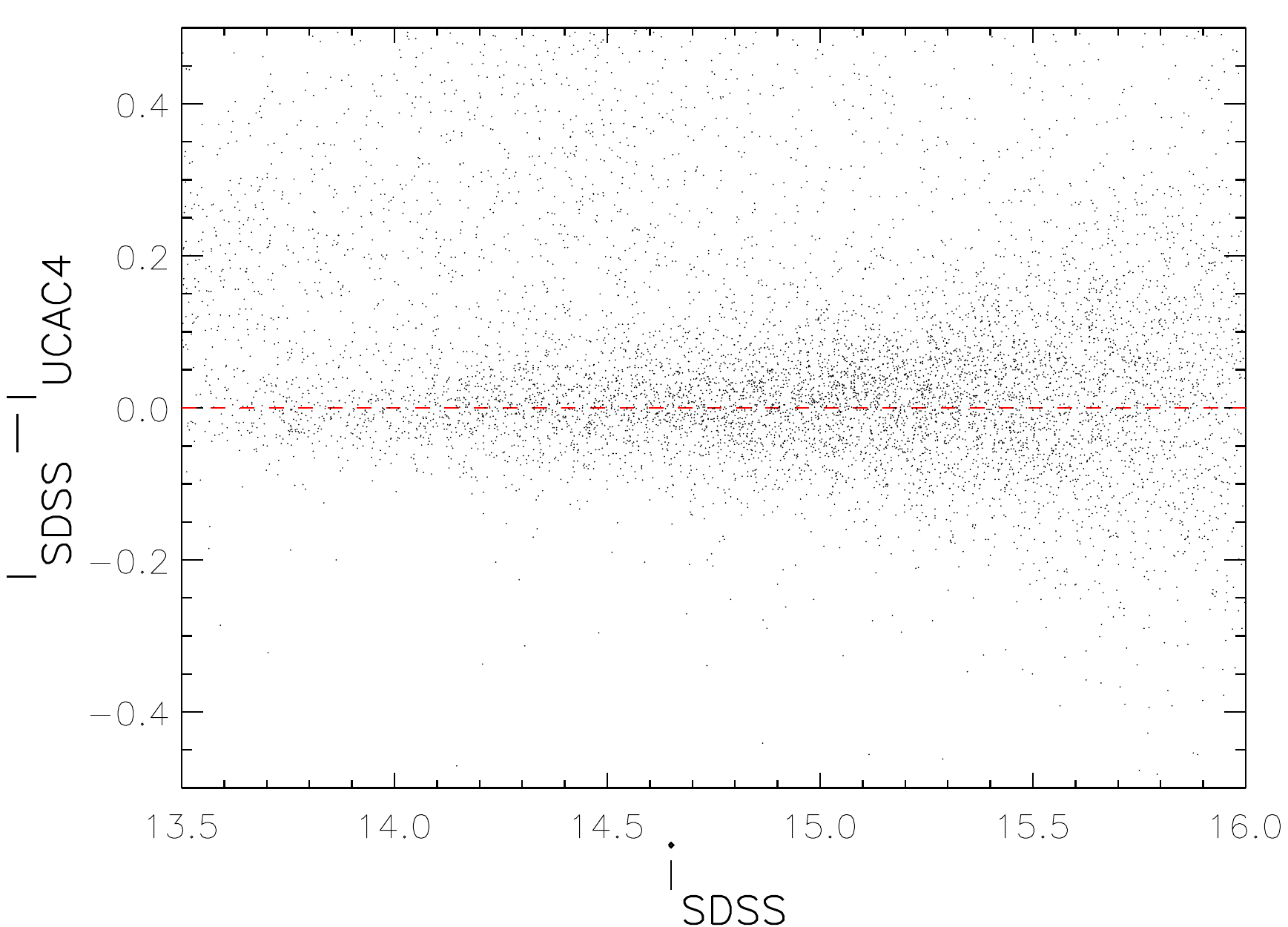}} 
	\end{tabular}
	\caption{Residual between the SDSS and UCAC4 photometries as a function of the SDSS magnitudes in the $r$ and $i$-bands (left and right panels, respectively).}
	\label{fig:SDSS_UCAC4}
\end{figure*}

The $r$ and $i$-band photometry in UCAC4 came from the AAVSO \footnote{\url{https://www.aavso.org}} Photometric All-Sky Survey \citep[][]{Henden2016}. These data were taken using the $r'$ and $i'$-band filters from SDSS, whose magnitudes are on the AB system and are close to the $r$ and $i$ magnitudes of SDSS\footnote{\url{http://www.sdss3.org/dr8/algorithms/fluxcal.php\#SDSStoAB}}. In Figure \ref{fig:SDSS_UCAC4} we show the residuals between the $r$ and $i$ magnitudes from SDSS and UCAC4 as a function of the SDSS magnitudes. We did not consider the sources having $>90$ per cent probability of being variables according to the CVSO and only worked with sources having photometric errors lesser than 0.05 mag. In average, these residuals are basically zero for sources brighter than the SDSS saturation limit ($\sim$14 mag), which indicates that the $r$ and $i$-band photometries from UCAC4 can be consider to be in the SDSS photometric system.

Thus, we worked with the following transformations from \citet{Jordi2006}, which use the $r$ and $i$-band magnitudes from SDSS:

\begin{equation} \label{eq:UCAC_1}
	R_c-r = -0.153*(r-i) - 0.117
\end{equation}
\begin{equation} \label{eq:UCAC_2}
	R_c-I_c = 0.930*(r-i) + 0.259
\end{equation}

Subtracting Transformation \ref{eq:UCAC_2} from Transformation \ref{eq:UCAC_1}:

\begin{equation} \label{eq:UCAC_3}
	I_c-r = -1.083*(r-i) -0.376
\end{equation}

\begin{figure*}
	\centering
	\begin{tabular}{cc}
		\subfloat{\includegraphics[width=.40\linewidth]{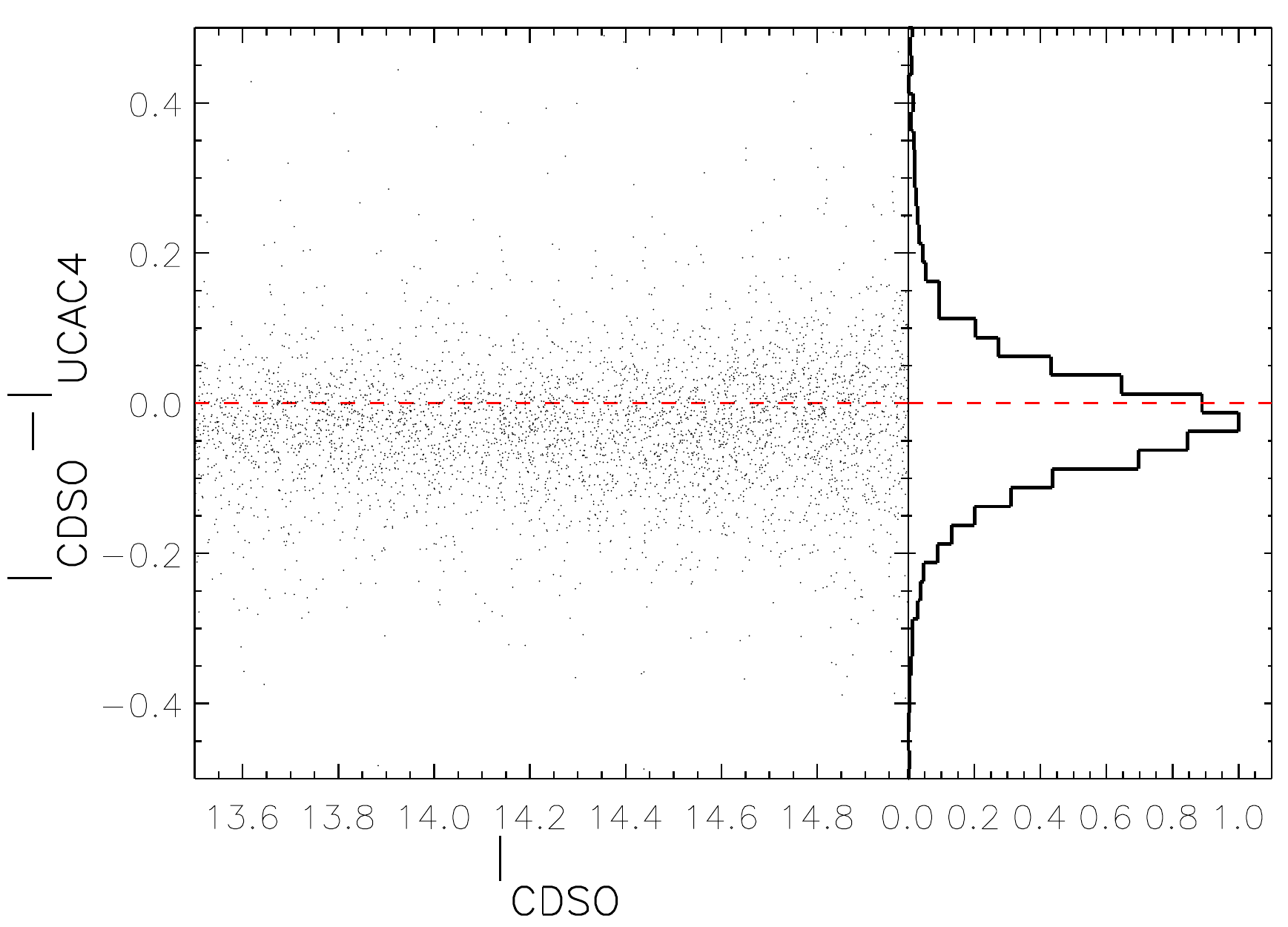}} & 
		\subfloat{\includegraphics[width=.40\linewidth]{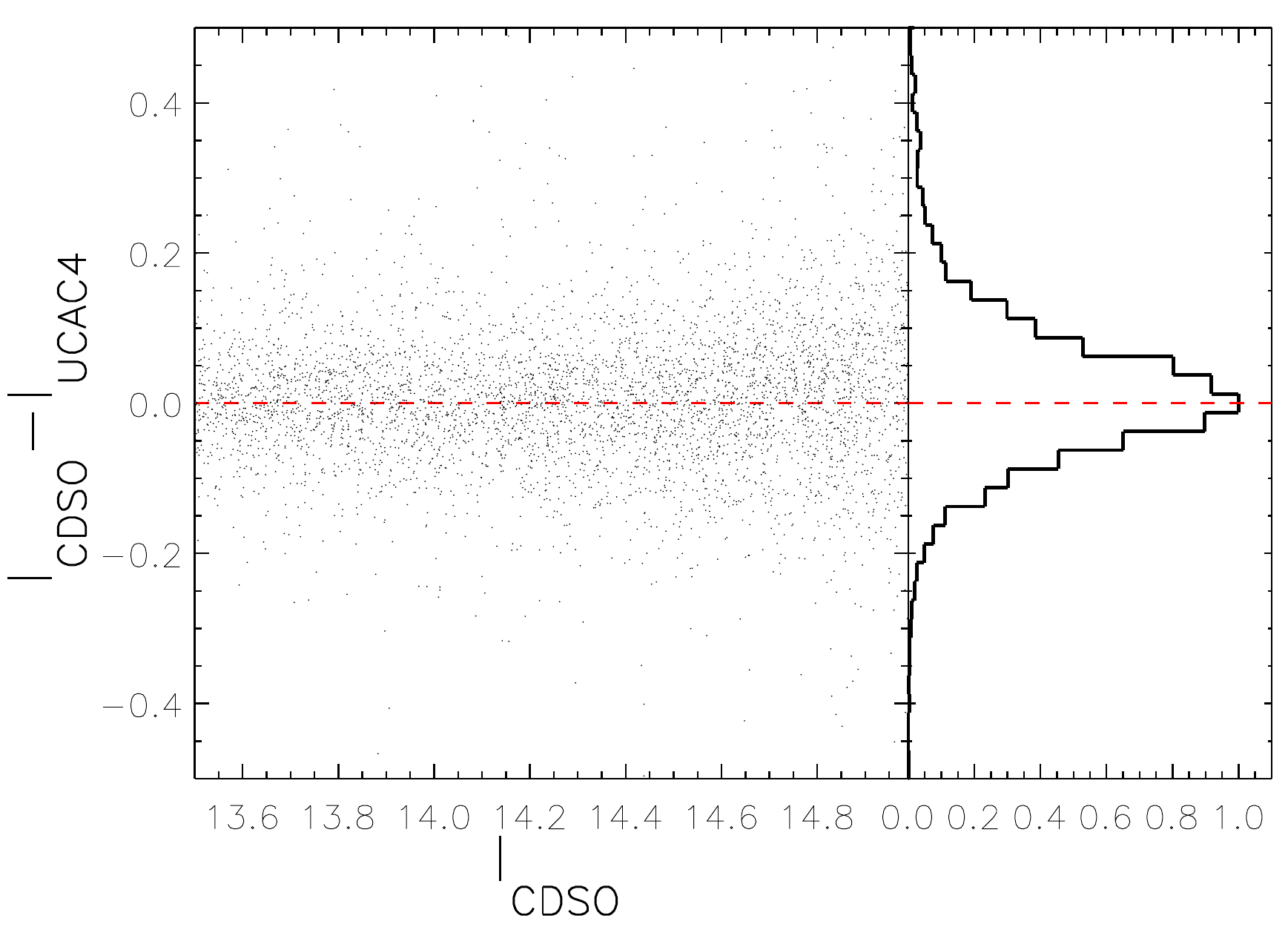}} 
	\end{tabular}
	\caption{$I_c$-residuals between the CDSO and UCAC4 after applying Transformation \ref{eq:UCAC_3} \citep[left panel; ][]{Jordi2006} and Transformation \ref{eq:UCAC_4} (right panel), which is a slight modification of Transformation \ref{eq:UCAC_3}.}
	\label{fig:trasformation_UCAC4}
\end{figure*}

We used Transformation \ref{eq:UCAC_3} to obtain the $I_c$ magnitudes considering the $r$ and $i$-band photometry from UCAC4. We compared the resultant $I_c$ magnitudes with those from the CDSO, which are already in the Cousin system. In the left panel of Figure \ref{fig:trasformation_UCAC4} we show the residual between the $I_c$ magnitudes from the CDSO and UCAC4, where we can see that the peak of the residual distribution is somewhat deviated from zero. Therefore, we did slight modifications to the coefficients of Transformation \ref{eq:UCAC_3} to have average residuals closer to zero. The resulting transformation is:

\begin{equation} \label{eq:UCAC_4}
	I_c-r = -1.323*(r-i) -0.353
\end{equation}

In the right panel of Figure \ref{fig:trasformation_UCAC4} we show the $I_c$ residuals between the CDSO and UCAC4 photometries after applying Transformation \ref{eq:UCAC_4} to the UCAC4 data. The peak of the $I_c$ residual histograms are essentially zero, with a RMS of 0.07 mag for all the sources within the CDSO saturation limit and the UCAC4 completeness limit (13-14.75 mag).

\subsection{DECam Data}
\begin{figure}
	\centering\includegraphics[width=0.40\textwidth]{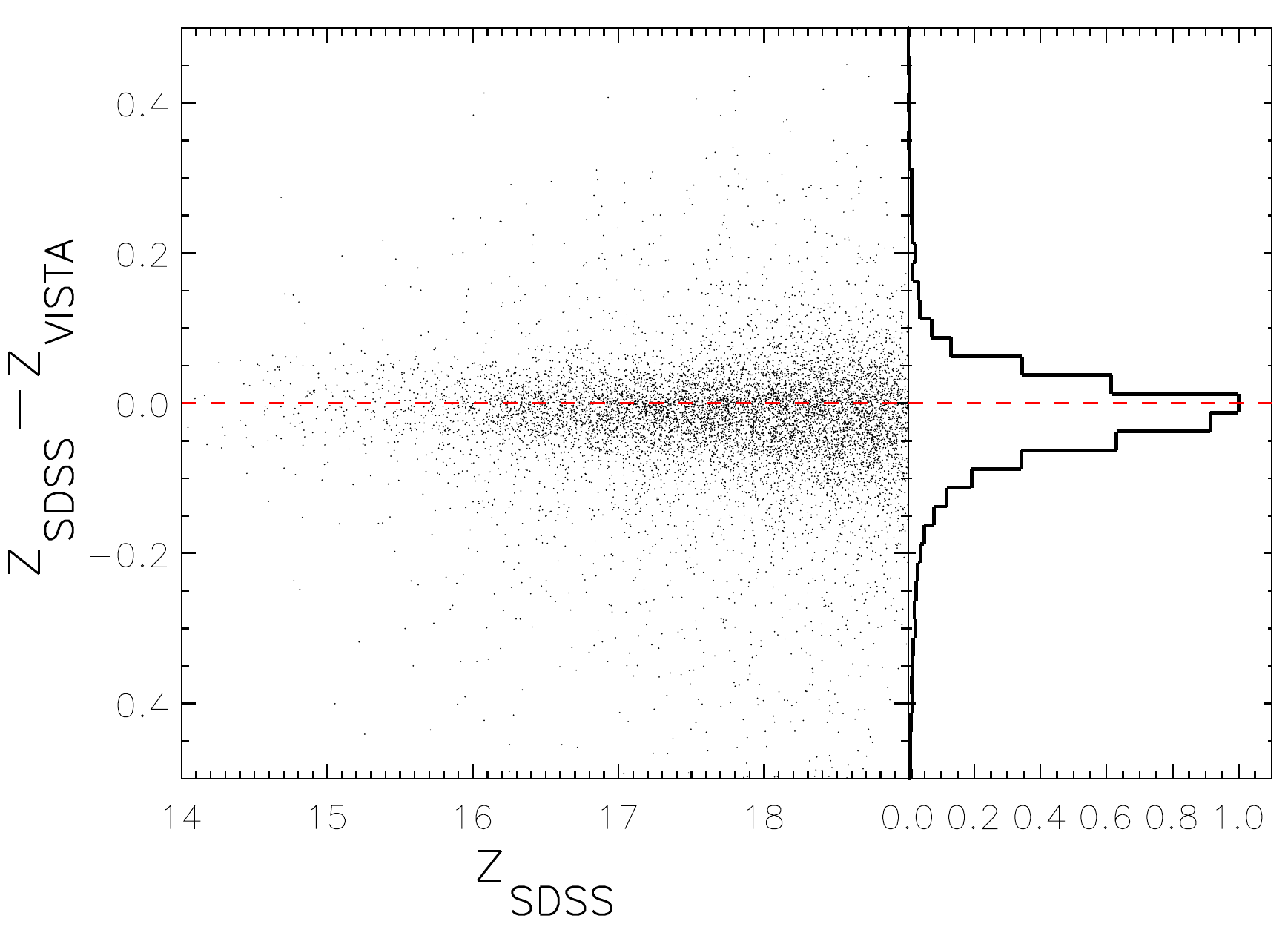}
	\caption{Residuals between the $z$ magnitudes in the SDSS system directly from the SDSS catalogue and from VISTA.}
	\label{fig:trasformation_VISTA}
\end{figure}

The $i$ filter used in our DECam observations is similar to the $i$ filter from SDSS (NOAO Data Handbook\footnote{\url{http://ast.noao.edu/sites/default/files/NOAO\_DHB\_v2.2.pdf}}). However, there is a colour dependence to transform the DECam data to the SDSS system. As we only have DECam photometry taken with the $i$ filter, in addition to these data we worked with the $Z$-band photometry from VISTA. This way, we will transform the DECam photometry only for the sources with VISTA counterpart, which is not an issue because for the selection of member candidates we used both catalogues. The $Z$-band photometry from VISTA is in the Vega system and to convert it to $z'$-band magnitudes in the AB system it is necessary to add the zero-point of 0.58 mag \citep{Pickles2010}. These $z'$-band magnitudes are not exactly the same as the $z$-band magnitudes in the SDSS system, there is a small shift of 0.02 mag which should be subtracted\footnote{\url{http://www.sdss3.org/dr8/algorithms/fluxcal.php\#SDSStoAB}}. Therefore, we added 0.56 mag to the $Z$-band photometry from VISTA to obtain the $z$-band magnitudes in the SDSS system. In Figure \ref{fig:trasformation_VISTA} we show the residuals between the $z$ magnitudes directly from SDSS and from VISTA after the addition of the offset. We removed the sources with $>90$ per cent probability of being variable according to the CVSO catalogue and we only considered sources with errors lesser than 0.05 mag. The average of the resultant residuals is -0.008 mag with a RMS of 0.04 mag.

\begin{figure*}
	\centering
	\begin{tabular}{cc}
		\subfloat{\includegraphics[width=.40\linewidth]{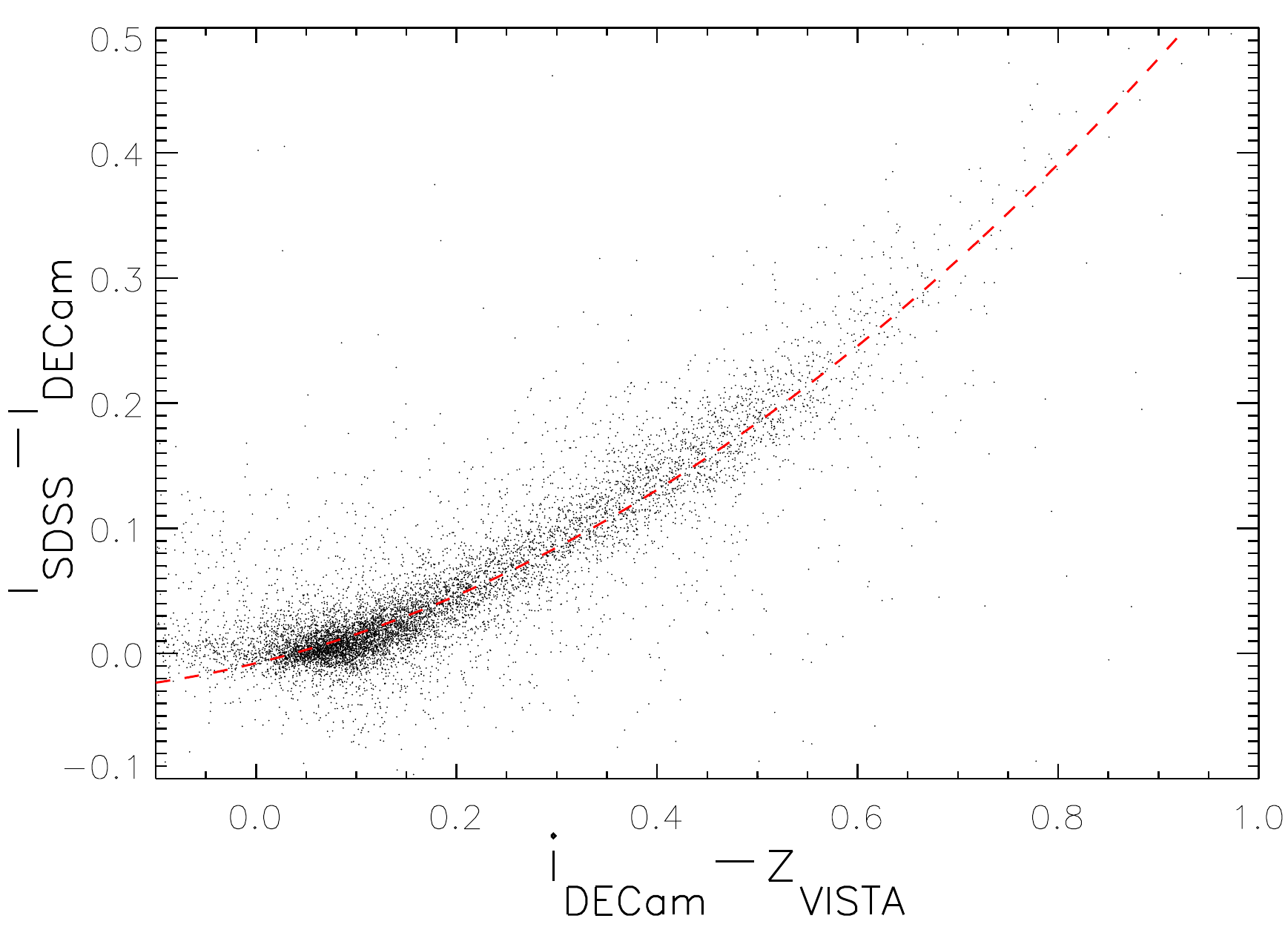}} & 
		\subfloat{\includegraphics[width=.40\linewidth]{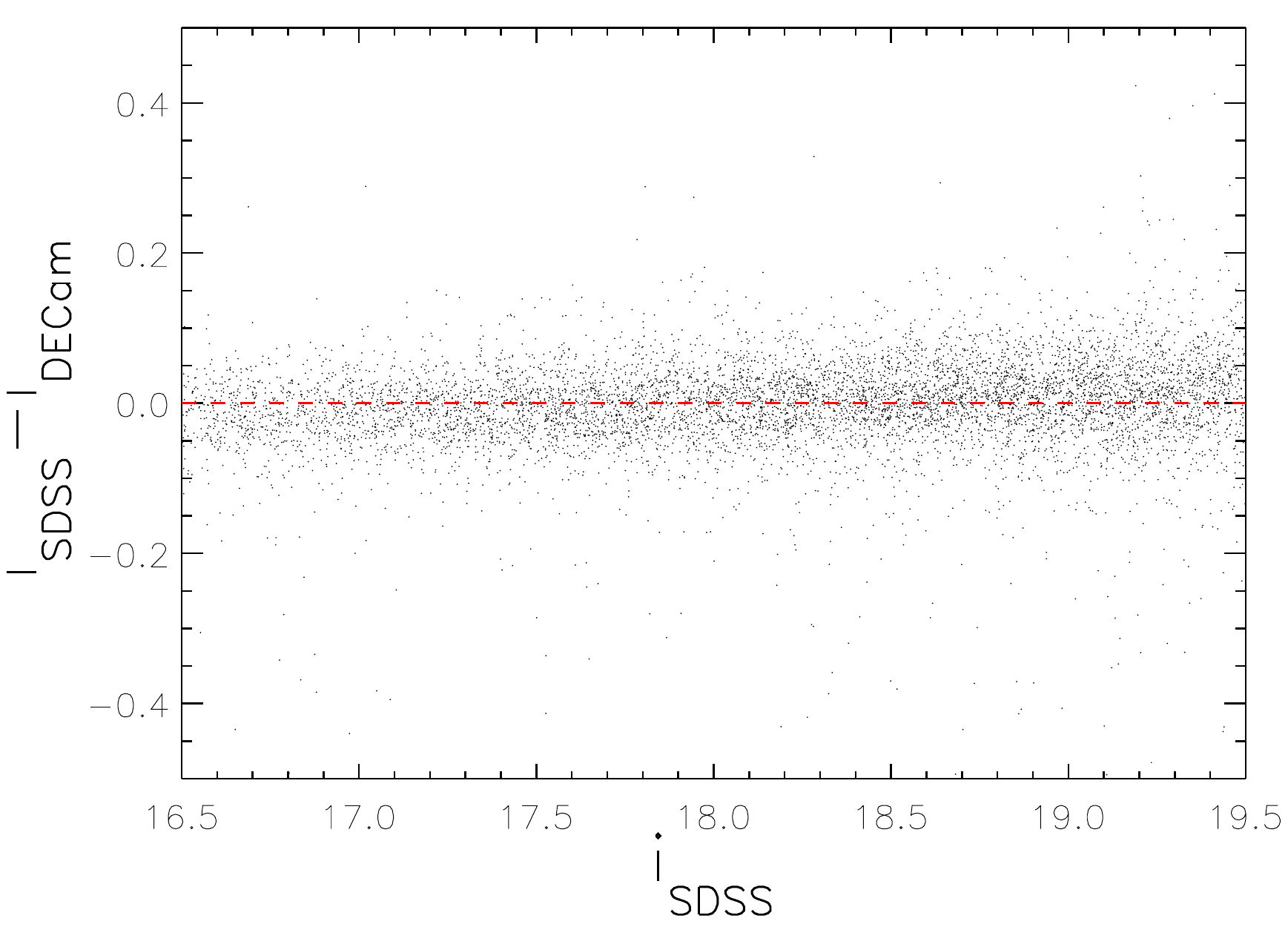}} 
	\end{tabular}
	\caption{{\bf Left panel:} Residuals between the $i$ magnitudes from the SDSS catalogue and from our calibrated DECam data as a function of the $i-z$ colour from DECam and VISTA data in the SDSS system. The red dashed line indicate the second order function fitted to the residuals. {\bf Right panel:} Residuals between the $i$-band photometries in the SDSS system directly from SDSS and from DECam after applying Transformation \ref{eq:DECam_SDSS}.}
	\label{fig:trasformation_DECam_SDSS}
\end{figure*}

In left panel of Figure \ref{fig:trasformation_DECam_SDSS} we show the colour dependence of the residuals between our calibrated DECam data and those from SDSS as a function of the $i-z$ colour combining the calibrated photometry from DECam and the photometry from VISTA converted to the SDSS system. The second order function that best fits the residuals is:

\begin{equation} \label{eq:DECam_SDSS}
	i-i_{DECam} = -0.008+ 0.194*(i_{DECam}-z)+ 0.381*(i_{DECam}-z)^2
\end{equation}

where $i_{DECam}$ are in the DECam system and $i$ and $z$ in the SDSS system.

We used Transformation \ref{eq:DECam_SDSS} to convert our calibrated DECam photometry to the SDSS system. In right panel of Figure \ref{fig:trasformation_DECam_SDSS} we show the residuals between the $i$-band magnitudes in the SDSS system obtained directly from SDSS and from our calibrated DECam data. The average of the residuals is 0.002 mag with a RMS of 0.04 mag.

\begin{figure*}
	\centering
	\begin{tabular}{cc}
		\subfloat{\includegraphics[width=.40\linewidth]{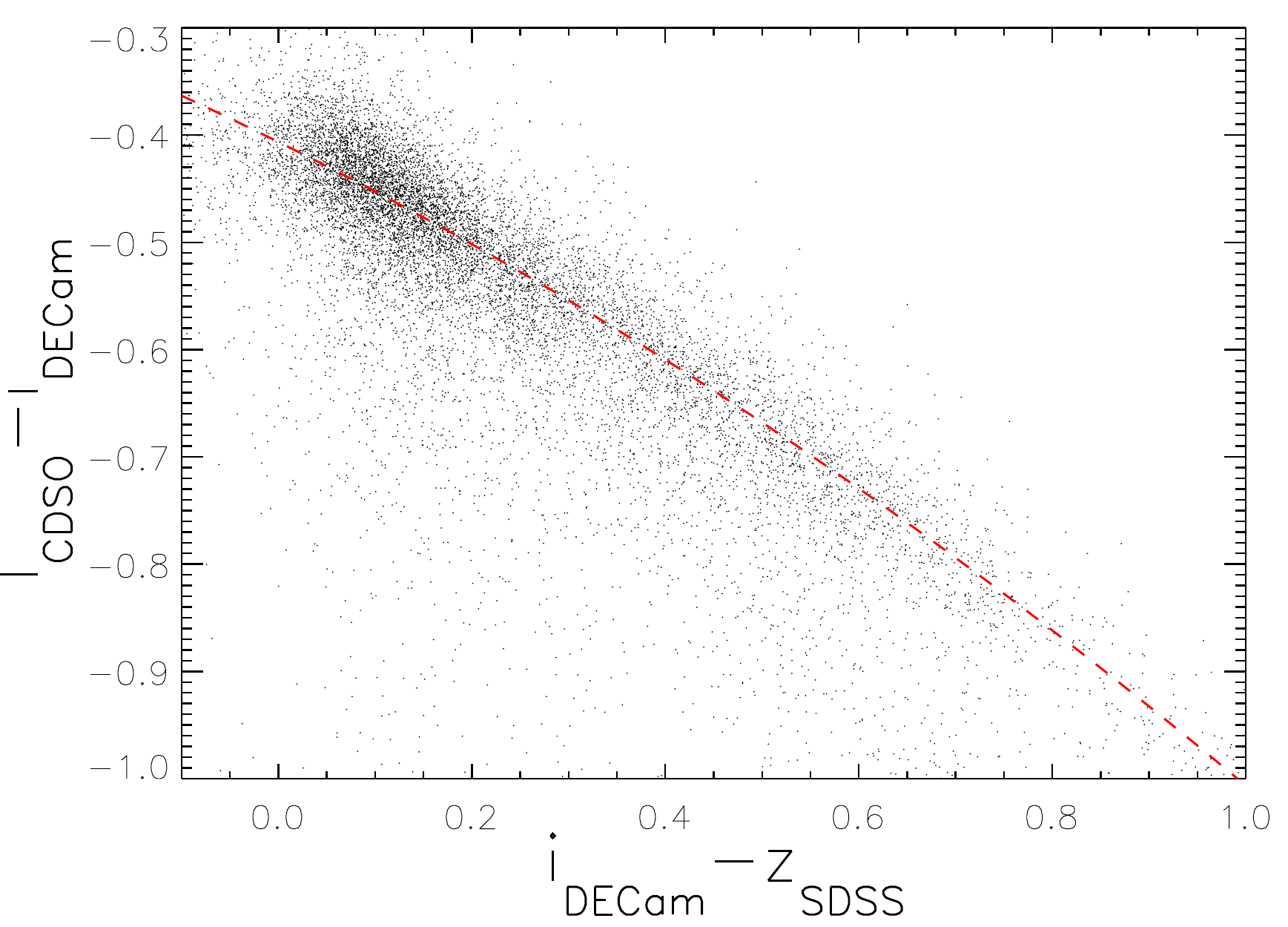}} & 
		\subfloat{\includegraphics[width=.40\linewidth]{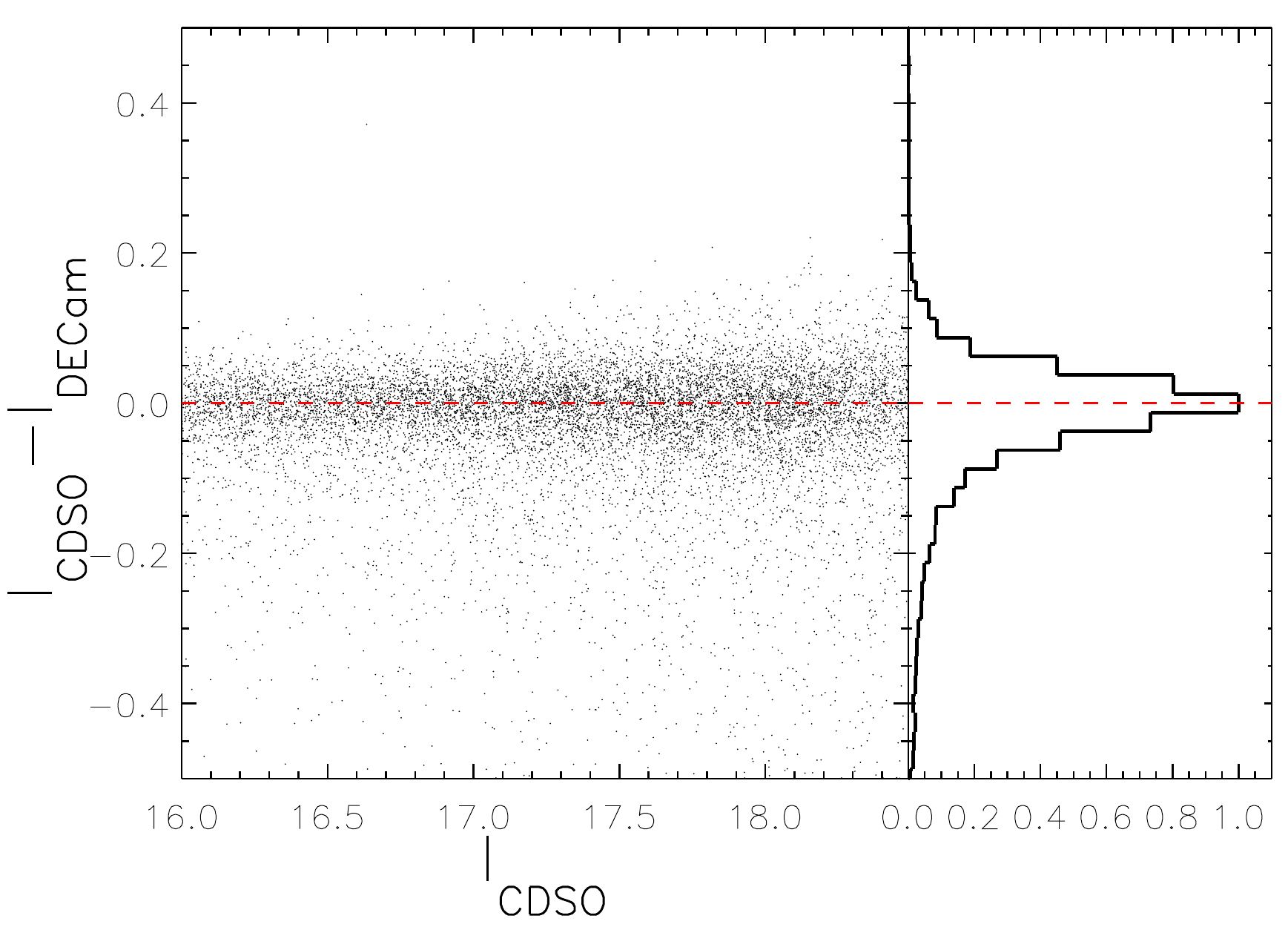}} 
	\end{tabular}
	\caption{{\bf Left panel:} Residuals between the $I_c$ magnitudes from the CDSO and the $i$ magnitudes from DECam as a function of the $i-z$ colour from DECam and VISTA data in the SDSS system. {\bf Right panel: } Residuals between the $I_c$-band photometries from the CDSO and DECam after applying Transformation \ref{eq:DECam_CDSO}.}
	\label{fig:trasformation_DECam_CDSO}
\end{figure*}

Finally, once we have both the $i$-band photometry from DECam and the $Z$-band photometry from VISTA in the SDSS system, we converted them to $I_c$ magnitudes in the Cousins system. In left panel of Figure \ref{fig:trasformation_DECam_CDSO} we show the $i-z$ dependence of the residual between the $I_c$ magnitudes from the CDSO survey and our DECam data in the SDSS system. The second order function fitted to the residual is:

\begin{equation} \label{eq:DECam_CDSO}
	I_c-i = -0.406-0.446*(i-z)-0.154*(i-z)^2
\end{equation}

where $I_c$ is in the Cousins system and $i$ and $z$ the SDSS system.

We used Transformation \ref{eq:DECam_CDSO} to obtain the $I_c$ magnitudes from our DECam and VISTA photometries in the SDSS system. In right panel of Figure \ref{fig:trasformation_DECam_CDSO} we show the residuals between the $I_c$ magnitudes from the CDSO and those obtained from our DECam data. We did not consider neither the sources having $>90$ per cent probability of being variable stars in the CVSO catalogue nor the sources with errors larger than 0.05 mag. The resultant residuals have an average of -0.001 mag with a RMS of 0.04 mag.

\section{Distance, Extinction and Proper Motion of 25 Ori}
\label{sec_app:25Ori_params}

\subsection{25 Ori Distance}
\label{sec_app:distance}
To estimate the 25 Ori distance we first compiled a list of 334 unique spectroscopically confirmed members of 25 Ori by \citet{Briceno2005,Briceno2007,Downes2014,Downes2015,Suarez2017,Briceno2019}. Then, we cross-matched this list with Gaia DR2 and with the BJ18 catalogue. 91 per cent of the confirmed members have Gaia DR2 parallaxes with uncertainties of $\le 20$ per cent. Using these parallaxes and the TOPCAT tool \citep{Taylor2005} with the method implemented by \citet{Bailer-Jones2015} and BJ18, we calculated the best distance estimates and the 25th and 75th percentile confidence intervals using the Exponentially Decreasing Space Density prior with a lenght scale of 500 pc. We consider these distance estimates as the distances of the compiled members and the percentiles as the uncertainties, which are very consistent with those obtained working with the inverse of the parallax as well as with the BJ18 distances.

In the left panel of Figure \ref{fig:cum_dist} we show the cumulative distribution of the distances of the confirmed members with parallax errors of $\le 20$ per cent, which cover a range from 127 to 545 pc, but there is a clear concentration of members around the 25 Ori expected distance with 94 per cent of them between 250 and 450 pc. From these distances we obtained that 25 Ori is 356$\pm$47 pc away, which is consistent with previous studies \citep{Briceno2007,Downes2014,Suarez2017,Briceno2019,Kounkel2018}.

\subsection{25 Ori Extinction}
\label{sec_app:extinction}
\begin{figure*}
	\centering
	\begin{tabular}{cc}
		\subfloat{\includegraphics[width=.40\linewidth]{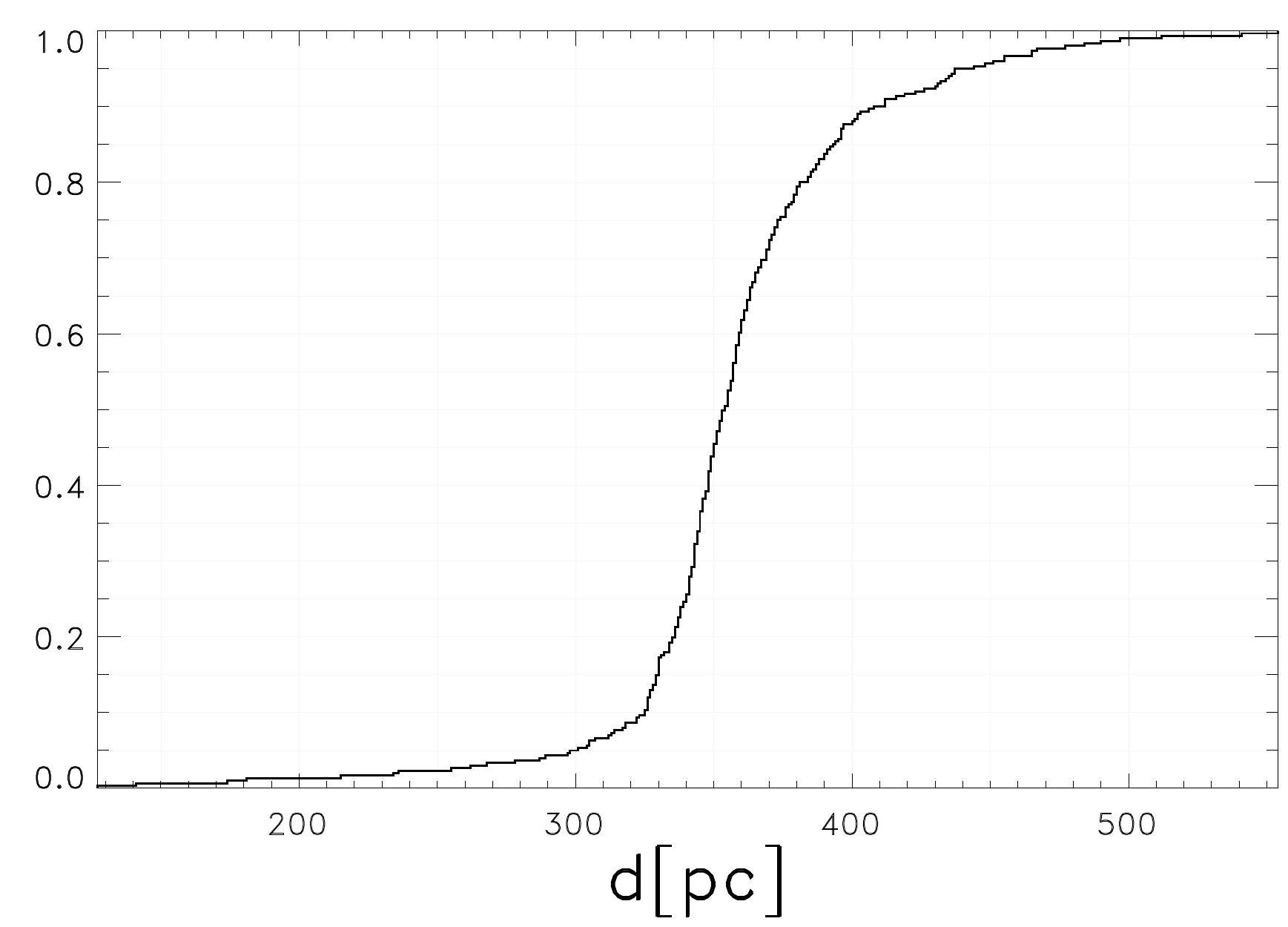}} & 
		\subfloat{\includegraphics[width=.40\linewidth]{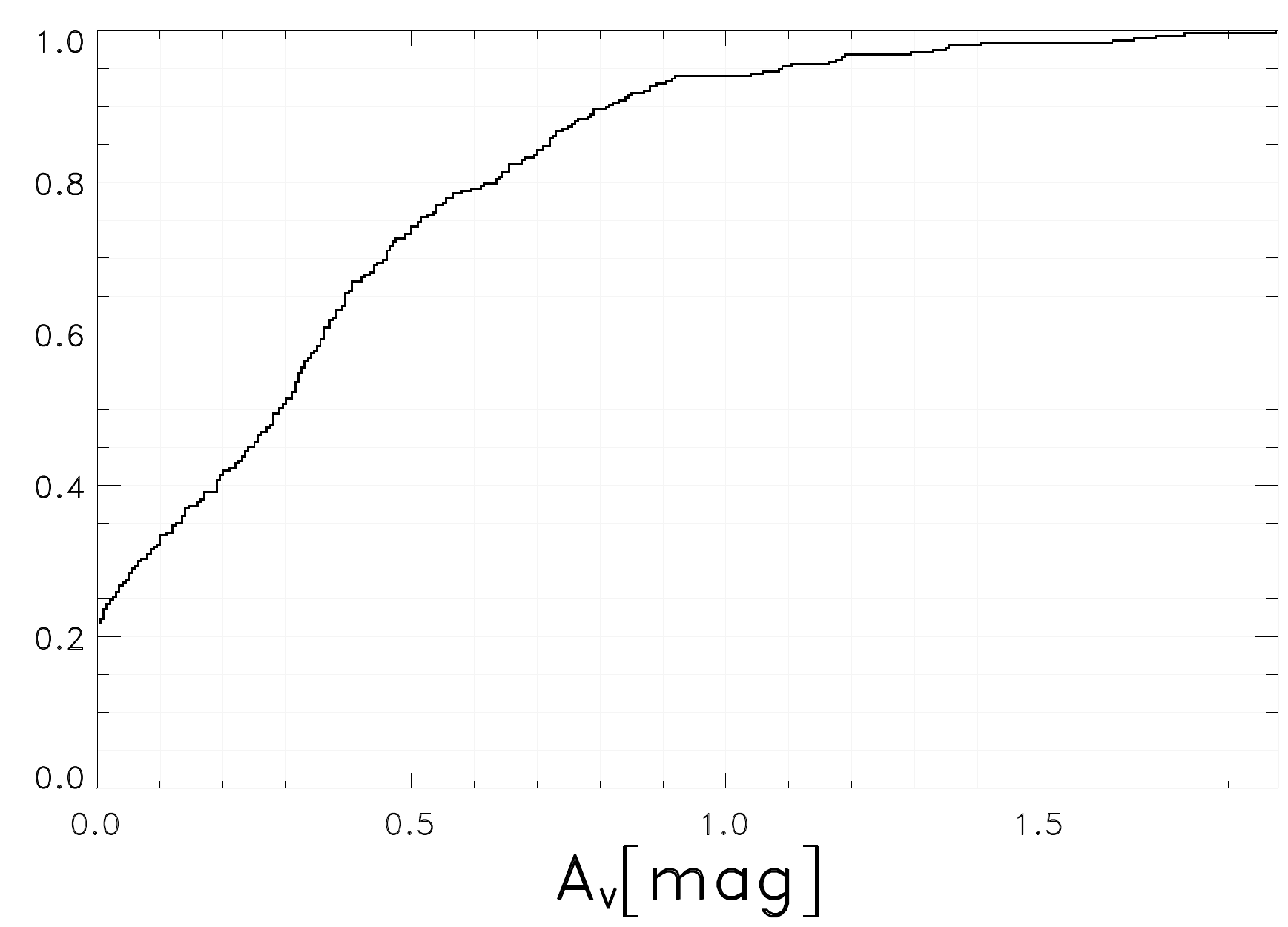}} 
	\end{tabular}
	\caption{Normalized cumulative distributions of the distances (left panel) and visual extinctions (right panel) for the spectroscopically confirmed members of 25 Ori by \citet[][]{Briceno2005,Briceno2007,Downes2014,Downes2015,Suarez2017,Briceno2019}. The distances are from the Gaia DR2 parallaxes with uncertainties of $\le 20$ per cent. The extinctions were mostly estimated through spectral analysis and combining optical and NIR photometry.}
	\label{fig:cum_dist}
\end{figure*}

About 96 per cent of the 334 confirmed members of 25 Ori by \citet{Briceno2005,Briceno2007,Downes2014,Downes2015,Suarez2017,Briceno2019} have reported visual extinctions obtained through spectroscopic analysis. In the right panel of Figure \ref{fig:cum_dist} we show the cumulative distribution of these extinctions, which go up to 1.88 mag (excluding two members with values of 3.53 and 6.29 mag) but more than 93 per cent of the members with reported extinction have values lower than 1 mag. Considering values up to 1.88 mag, the mean visual extinction of 25 Ori is 0.35$\pm$0.35 mag. If we consider values lower than 1 mag, the 25 Ori mean visual extinction is 0.29$\pm$0.26 mag. As expected, both values are consistent with previous studies \citep{Kharchenko2005,Briceno2005,Briceno2007,Downes2014,Suarez2017,Briceno2019}. 

\subsection{25 Ori Proper Motion}
\label{sec_app:proper_motion}
To estimate the proper motion of 25 Ori we used the list of 25 Ori confirmed members \citep{Briceno2005,Briceno2007,Downes2014,Downes2015,Suarez2017,Briceno2019} having Gaia DR2 parallaxes with errors of $\le 20$ per cent. We discarded 6 members with clearly discrepant proper motions and 17 members forming a possible distinct overdensity (perhaps ASCC 18). With the remaining 81 per cent of the sample we estimated that the mean proper motion of 25 Ori is $\mu_\alpha = 1.33\pm0.46$ mas yr$^{-1}$ and $\mu_\delta = -0.23\pm0.55$ mas yr$^{-1}$.

\section{Distances and Extinctions for the Member Candidates and Contaminants}
\label{sec_app:distance_extinction}
As we do not have distances and extinctions for all the member candidates (86 per cent have distances and 18 per cent have visual extinctions) and contaminants, we need to assign these values to the whole samples to have consistency with the 25 Ori members. The most common way to do this in photometric studies in the literature is to consider the mean distance and extinction of the cluster for all the member candidates. Here, we can take advantage of the Gaia DR2 parallaxes as well as of the previous spectroscopic studies in 25 Ori to use a statistically more robust technique. Considering the inversion of the normalized cumulative distribution of the distances of the 25 Ori confirmed members (left panel of Figure \ref{fig:cum_dist}), we created random realizations to assign distance values to all our member candidates and contaminants. We also assigned extinction values to these samples in a similar way, but considering the normalized cumulative distribution of the reported visual extinctions of the 25 Ori confirmed members (right panel of Figure \ref{fig:cum_dist}). This way, the distance and extinction values we assigned to each candidate and contaminant are consistent with those for the confirmed members of 25 Ori.

\end{document}